\documentclass[a4paper,11pt]{article}
\usepackage[utf8]{inputenc}
\pdfoutput=1
\usepackage{jheppub,float}
\usepackage[normalem]{ulem}
\usepackage{xcolor}
\usepackage{amsmath}
\usepackage{framed}
\usepackage{varioref}
\colorlet{shadecolor}{lightgray}
\usepackage{hyperref}
\usepackage{mathrsfs}
\usepackage{parskip}
\labelformat{equation}{Eq.~(#1)}

\newcommand{\n}{\nabla}

\title{\boldmath Soft Photon theorem in the small negative cosmological constant limit}

\author[a]{Nabamita Banerjee,}
\author[b]{Karan Fernandes,}
\author[a]{and Arpita Mitra.}

\affiliation[a]{Indian Institute of Science Education \& Research Bhopal,\\
	Bhopal Bypass Road, Bhauri, Bhopal 420 066,\\ Madhya Pradesh, India.}
\affiliation[b]{Harish-Chandra Research Institute,\\ Chhatnag Road, Jhusi, Prayagraj 211019,\\ Uttar Pradesh, India}
\emailAdd{nabamita@iiserb.ac.in, karanfernandes@hri.res.in, arpitam@iiserb.ac.in}
\abstract{We study the effect of electromagnetic interactions on the classical soft theorems on an asymptotically AdS background in 4 spacetime dimensions, in the limit of a small cosmological constant or equivalently a large AdS radius $l$. This identifies $1/l^2$ perturbative corrections to the known asymptotically flat spacetime leading and subleading soft factors. Our analysis is only valid to leading order in $1/l^2$. The leading soft factor can be expected to be universal and holds beyond tree level. This allows us to derive a $1/l^2$ corrected Ward identity, following the known equivalence between large gauge Ward identities and soft theorems in asymptotically flat spacetimes.}

\begin{document} 
\maketitle
\flushbottom

\section{Introduction}

Soft theorems are statements about quantum scattering amplitudes when one or more of the external particles go soft, i.e. their momenta $k^{\mu} \rightarrow 0.$ These theorems state that a $(m+n)$ point scattering amplitude ${\cal{A}}_{m+n}$, where $m$ number of external particles go soft, is proportional to ${\cal{A}}_{n}$, the $n$ point scattering amplitude involving the other hard particles. The proportionality factor is universal at leading order, irrespective of the details of the interactions \cite{Weinberg:1964ew, Weinberg:1965nx}. The soft factor is also divergent at leading order in the soft momenta expansion and depends on properties of the hard particles. The theorems are valid for any gauge invariant quantum field theory in any spacetime dimensions. In particular, the soft photon theorem follows from the U(1) gauge invariance \cite{Kapec:2014zla, He:2014cra} and the soft graviton theorem follows from the diffeomorphism invariance of quantum field theories \cite{Strominger:2013jfa, Cachazo:2014fwa, Campiglia:2015yka}. Recent works \cite{Lysov:2014csa, Schwab:2014xua, Campiglia:2014yka, Casali:2014xpa,Broedel:2014fsa, Conde:2016csj, Chakrabarti:2017zmh, Chakrabarti:2017ltl, Laddha:2017vfh, AtulBhatkar:2018kfi, Addazi:2019mjh,Sahoo:2020ryf} have extended soft theorems beyond leading orders, with results for subleading and sub-subleading soft theorems in gravitational and U(1) gauge theories. In four spacetime dimensions there exists an additional subtlety, in that the subleading soft factor diverges as the logarithm of the frequency due to the existence of asymptotically non-vanishing long range interactions. To be precise, below we write the exact statement of the single soft photon theorem to subleading order in four spacetime dimensions \cite{Sahoo:2018lxl}:

\begin{align}
S_{\rm em}^{\text{flat}} &= S_{\rm em; \text{leading}}^{\text{flat}} + S_{\rm em; \text{subleading}}^{\text{flat}} \,, \qquad \text{with} \notag\\
S_{\rm em; \text{leading}}^{\text{flat}} &= \sum_{a=1}^n  q_{(a)} \frac{\epsilon_{\mu} p^{\mu}_{(a)}}{p_{(a)}.k} \,, \label{sphff2}\\
S_{\rm em; \text{subleading}}^{\text{flat}} &= i \sum_{a=1}^n  q_{(a)}\frac{\epsilon_{\nu} k_\rho j^{\rho \nu}_{(a)}}{p_{(a)}.k}\notag\\&=i \ln \omega^{-1}\sum_{a=1}^n  q_{(a)} \frac{\epsilon_{\nu} k_\rho \bigg(c_{(a)}^{\rho}p_{(a)}^{\nu}-c_{(a)}^{\nu}p_{(a)}^{\rho}\bigg)}{p_{(a)}.k} + \cdots
\label{sphff.sub2}
\end{align}
where $k^{\mu }$ and $\epsilon_{\mu}$ are respectively the momentum and polarization of the soft photon, while $q_{(a)}$, ${p}^{\mu}_{(a)}$ and ${j}^{\mu \nu}_{(a)}$ are the charges, asymptotic momenta and angular momenta of the $n$ hard particles. In going from the first to second equality of \ref{sphff.sub2}, we have an expansion in terms of the classical trajectories of the hard particles. The individual trajectories involve logarithmic contributions in four dimensions due to the presence of long range interactions of the electromagnetic fields. We can expand the classical trajectories of these particles to find the following leading order contribution in proper time $$r^{\mu}_{(a)}(t) = \eta_{(a)} \frac{p_{(a)}^{\mu}}{m_{(a)}}t + c^{\mu}_{(a)} \ln \vert t \vert + \cdots,$$ where $t$ is the proper time along the particle trajectories, $\eta_{(a)}$ is $+1$ for incoming particles and $-1$ for outgoing particles, $m_{(a)}$ are the particle masses and $c^{\mu}_{(a)}$ are coefficients which depend on the long range electromagnetic force. Using this expression for $r^{\mu}_{(a)}(t)$ in $j^{\mu \nu}_{(a)} = r^{\mu}_{(a)}(t)p_{(a)}^{\nu} - r^{\nu}(t)p_{(a)}^{\mu}$ then provides the term in the second line of \ref{sphff.sub2} on replacing $t$ with $\omega^{-1}$. The additional terms not described in the second line of \ref{sphff.sub2} are quantum corrections. These terms can be ignored as long as they are much smaller than the classical scattering contribution. This will be the case when the wavelength of the soft particles are much larger than the impact parameter involved in the scattering and when the total radiated energy is less than the energy of the
scatterer. In such cases, we can derive the universal contributions entirely from the low frequency limit of the (gauge invariant) classical radiative fields. This relation is provided by the classical soft photon theorem \cite{Sahoo:2018lxl, Laddha:2018rle, Laddha:2018myi},
\begin{align}
    \lim_{\omega \to 0}\epsilon^{\mu}\tilde{a}_{\mu}\left(\omega\,, \vec{x}\right) &= e^{i \omega R} \left(\frac{\omega}{2 \pi i R}\right)^{\frac{D-2}{2}} \frac{1}{2 \omega} S^{\text{flat}}_{\text{em}} \notag\\
   & = - \frac{i}{4 \pi R} e^{i\omega R} S^{\text{flat}}_{\text{em}} \qquad \qquad \text{for}\; D = 4\,,
    \label{class.softem}
\end{align}

where $\tilde{a}_{\mu}$ is the radiative component of the electromagnetic field in frequency space, $D$ is the spacetime dimension and $R$ denotes the distance of the soft photon from the scatterer. A similar analysis holds for the soft graviton theorem~\cite{Sahoo:2018lxl,Laddha:2018vbn,Laddha:2019yaj, Saha:2019tub}. 

An interesting question to ask is: how does the above story change when we study a quantum field theory in an asymptotically non-flat background? Since the gauge invariance remains intact even for asymptotically non-flat theories, we expect a version of soft theorems to be valid in this case as well. Non-asymptotically flat backgrounds, particularly of the kind of asymptotically Anti-de Sitter (AdS) or de Sitter (dS) types are of great importance in physics. AdS arises as an interesting gravity background for some exact computations in the context of String Theory and AdS/CFT conjectures. On the other hand dS spacetime has its importance in cosmology. Thus, understanding aspects of soft theorems in these spacetimes are important. In this paper we shall look for classical soft photon theorems in asymptotically AdS spacetime, where the radius of the AdS space is considered to be large (we shall make this condition more precise in later sections). Our results, with slight modifications, are also valid for asymptotically dS spacetimes in the large radius limit.  
 
AdS (dS) is a solution of Einstein's gravity with a negative (positive) cosmological constant.
AdS spacetimes have an effective potential under which particles behave like being confined in a box. The null rays bounce back from the timelike boundary an infinite number of times. This creates the main obstacle in defining the usual ``in" and ``out" states for a quantum field theory in AdS backgrounds. Thus, unlike in asymptotically flat theories, the definition of the usual scattering amplitudes \cite{Gary:2009mi,Penedones:2010ue,Fitzpatrick:2011ia, Rastelli:2016nze} and hence a soft theorem is not known for quantum field theories defined in an asymptotically AdS spacetime\footnote{a related work can be found in \cite{Hijano:2020szl}}. Instead we look for a possible soft factorization on taking the classical limit of scattering amplitudes in AdS backgrounds. This gives us the analogue of classical soft theorems known from asymptotically flat spacetime classical scattering processes. While computing the classical radiation profile, for technical simplification, we consider the value of cosmological constant to be small , or equivalently the radius $l$ of AdS large and treat it as a perturbation parameter in our computations. Our results are exact up to order $1/l^2$ of the AdS radius. Physically, we think of studying a scattering process in an asymptotically flat theory modified by a small potential (inversely proportional to the square of the AdS radius). Thus our results provide us perturbative corrections to order $1/l^2$ of known results for classical photon and graviton radiation profiles in the asymptotically flat Reissner-Nordstr\"om case \cite{Fernandes:2020tsq}. The details of the scattering process we consider will be discussed in later sections.

Finally to study the ``soft limit" of the classical radiation in asymptotically AdS spacetime, we consider a double scaling limit \cite{Banerjee:2020dww}: where the frequency of radiation and cosmological constant simultaneously tend to zero, keeping their ratio finite. This is due to the fact that a radiation mode in a theory that asymptotes to AdS spacetime has a minimum frequency inversely proportional to the size of the AdS and hence the frequency of the radiation cannot limit to a zero value. In other words, there is a mass gap in AdS that restricts the usual soft limit. Physically the double scaling limit implies that we consider the radiation limits to a strictly soft one as the space is limiting to an asymptotically flat spacetime.  By taking this limit, we find the classical soft photon theorem to leading and sub-leading order in an asymptotically AdS theory, in the large AdS radius limit.

 On asymptotically flat spacetimes, Weinberg's soft theorems for scattering amplitudes are known to be equivalent to Ward identities for large gauge transformations \cite{He:2014cra},  \cite{Lysov:2014csa},\cite{He:2014laa, Hamada:2018vrw, Campiglia:2019wxe, AtulBhatkar:2019vcb}. These identities represent the soft charge conservation across null infinity $\mathscr{I}$. For asymptotically flat spacetimes, it is well known by now that the classical soft factor provide the same leading (quantum) soft factor in the classical limit up to the usual gauge ambiguity. Hence we can as well recast the classical soft theorem in terms of the large gauge  Ward identity. In our present study, as the classical soft photon theorem receives a $1/l^{2}$ correction (to its flat space form) in taking the $l \to \infty$ limit of AdS spacetimes, we expect the equivalence to imply a $1/l^{2}$ correction of the usual large gauge Ward identity. This physically amounts to deriving $1/l^{2}$ perturbative modifications in the Ward identity that correctly reproduces the modified classical soft theorem.
 
 The large gauge Ward identity on an asymptotically AdS spacetime involve two subtleties related to their derivation and definition on an asymptotic surface. A formal derivation of the large gauge Ward identity in AdS spacetimes is complicated by the fact that there does not appear to be a unique large $r$ saddle point corresponding to the low frequency result, unlike in asymptotically flat spacetimes. As mentioned, the $1/l^{2}$ corrections to the soft factor results from a double scaling limit on the cosmological constant and frequency, thus receiving contributions across different length scales. We can nevertheless infer this Ward identity from the soft photon theorem, following the procedure used in~\cite{He:2014cra} to demonstrate the equivalence. We find specific corrections of the soft photon mode and gauge parameter that provide a Ward identity equivalent to the classical soft photon theorem up to $1/l^{2}$ corrections. It has recently been demonstrated that conformal Ward identities are equivalent to the Weinberg's soft theorems defined on a flat spacetime patch resulting from the $l \to \infty$ limit on asymptotically AdS spacetimes~\cite{Hijano:2020szl}. Our Ward identity result has a natural interpretation of being defined at ``null infinity" on this patch near the center of AdS \cite{Hijano:2020szl, Compere:2019bua}, providing the leading $1/l^2$ corrections.

The current paper generalizes our previous work \cite{Banerjee:2020dww} on the effect of the small AdS potential on the classical soft graviton theorem by including an electromagnetic interaction. We also find the effect of the small AdS potential on the classical soft photon theorem. The paper is organised as follows: In section \ref{sec2}, we review basic properties of AdS Reissner-Nordstr\"om spacetime and then study its perturbations by introducing a charged point probe particle. In section \ref{sec3}, we have obtained the solution to the gauge and gravity radiations. Next in section \ref{sec4}, we study the soft limit and extract the classical soft photon factor from classical radiation profile. In taking a similar limit, in section \ref{sec5} we state the results for the classical soft graviton factor. It turns out that the charge of the central black hole, considered as the scatterer in the classical probe scattering process, has no explicit effect on the soft graviton factor. Finally in section \ref{sec6}, we find the Ward identity of large gauge transformations, perturbatively modified to $1/l^2$ order using our classical soft photon factors that we derived in section \ref{sec4}. We end the paper with a 
conclusion and some interesting open questions in section \ref{sec7}. Appendix \ref{SGT} contains the computation details for the classical soft graviton factor.

\section{Perturbations of AdS Reissner-Nordstr\"om spacetime}\label{sec2}
In this paper, we are interested in studying the classical soft photon theorem in asymptotically AdS backgrounds. To achieve this, we study the classical scattering of a charged and massive probe particle by a Reissner-Nordstr\"om black hole placed in an asymptotically AdS spacetime in 4 spacetime dimensions. The equations of motions result from the action, which consists of the Einstein-Hilbert term and the Maxwell term, 
\begin{equation}
S=\frac{1}{16\pi G}\int d^4 x \sqrt{-g}(R-2\Lambda)-\frac{1}{16\pi}\int d^4 x \sqrt{-g}~F_{\mu\nu}F^{\mu\nu}.\label{ac}
\end{equation}
In \ref{ac} $R$ is the Ricci scalar for metric $g_{\mu\nu}$, $\Lambda$ is the cosmological constant, $G$ is Newton's constant and $F_{\mu\nu}=A_{\nu,\mu}-A_{\mu,\nu}$ is the field strength tensor of the electromagnetic field $A_\mu$. {\footnote{The Maxwell action is written using Heaviside units and further details can be found in the Appendix E of \cite{Wald:1984rg}.}} We use the standard convention of denoting partial derivatives by subscripted commas and covariant derivatives with semi-colons. Varying the action in \ref{ac} with respect to the metric tensor one gets the Einstein equations,
\begin{equation}
R_{\mu\nu}-\frac{1}{2}R g_{\mu\nu}+\Lambda g_{\mu\nu}= 8\pi G T_{\mu\nu}^{EM}\label{emm},
\end{equation}
where 
\begin{align*}
T_{\mu\nu}^{EM} =\frac{1}{4\pi}\left(F_{\mu\alpha}F_{\nu\beta}g^{\alpha\beta}-\frac{1}{4}g_{\mu\nu}F_{\alpha\beta}F_{\gamma\delta}g^{\alpha\gamma}g^{\beta\delta}\right) \,. 
\end{align*}
Similarly for the gauge field $A_{\mu}$ we get the source-free Maxwell equations
\begin{equation}
\frac{\sqrt{-g}}{4\pi}F^{\mu\nu}{}_{;\nu}=0 \,.\label{max}
\end{equation}
The solutions of equations \ref{emm} and \ref{max} for a static spherically symmetric spacetime with mass M, charge Q and a negative cosmological constant $\Lambda=-3/l^2$, provide the metric 
\begin{equation}
ds^2=-f(r)dt^2+\frac{d r^2}{f(r)}+r^2(d\theta^2+\sin^2 \theta d\phi^2)\label{metric},
\end{equation}
with a gauge potential
\begin{equation}
    A_0=\frac{Q}{r}\,.
\end{equation}
The lapse function $f(r)$ in global coordinates takes the form
\begin{equation}
f(r)=1-\frac{2GM}{r}+\frac{GQ^2}{r^2}-\Lambda\frac{r^2}{3}=1-\frac{2GM}{r}+\frac{GQ^2}{r^2}+\frac{r^2}{l^2}.
\end{equation}
Since we are interested in studying the radiation emitted by the scattering of a probe particle moving in an unbounded trajectory on the spacetime (from the point of view of an asymptotic observer) we introduce isotropic coordinates. We refer the reader to \cite{Banerjee:2020dww} for further justification on choosing this particular coordinate system. In these coordinates, the resulting radiation will be isotropic in all spatial directions. We assume that the probe particle with mass $m\, (\ll M)$ and charge $q\, (\ll Q)$ has a large impact parameter from the black hole which implies $GM/r\ll 1$ and $\sqrt{G}Q/r\ll 1$. In addition, we also truncate our metric up to $1/l^2$ terms, as we consider radiative solutions in the large cosmological constant limit. Therefore our result will be valid in the regime $\sqrt{G} Q \leq G M\ll r\ll l$.\footnote{The equality of $\sqrt{G}Q$ and $G M$ holds in the extremal limit} 

The metric \ref{metric} in isotropic coordinates, expanded up to quadratic order in $\rho$, takes the form  
\begin{equation}
ds^2=-g_{00}dt^2+g_{ij}dx^i dx^j, \label{rn.met}
\end{equation}
where
\begin{align}
g_{00} = -\left(1 -  \frac{2 G M}{\rho} +  \frac{G Q^2}{\rho^2} + \frac{\rho^2}{l^2}\right) \,, \qquad g_{0i} &= 0 \,, \qquad g_{ij} = \delta_{ij} \left(1 +  \frac{2 G M}{\rho} -  \frac{G Q^2}{\rho^2} + \frac{\rho^2}{2 l^2}\right)\,
\label{rn.metb}.
\end{align}
Here $(i,j)=1,2,3$, run over spatial directions and $\rho=|\vec{x}|$. The isotropic coordinate $\rho$ is related to the Schwarzschild coordinate `$r$' by
\begin{equation}
    \rho=r\left(1-\frac{GM}{r}+\frac{GQ^2}{2r^2}-\frac{r^2}{4l^2}\right)\,.
\end{equation} 

We now impose the assumptions discussed above to express the metric of \ref{rn.metb} in a form relevant for our calculations. We set $8\pi G = 1$ in the following. Therefore we will replace $G$ by $1/8\pi$ in the remainder of the paper. The condition of a large impact parameter amounts to considering the leading order contribution of the gravitational and electromagnetic potential, which we will denote by $\phi \left(\vec{x}\right)$.
The potential goes like $r^{-1}$ and can be defined either with respect to the mass or the charge. The analysis in this paper is independent of either choice. We define

\begin{equation}
\phi\left(\vec{x}\right)= -\frac{M}{8 \pi \rho}\,.
\label{rn.pot}
\end{equation}

The gauge potential can then be expressed as

\begin{equation}
A_0(\vec{x}) = \frac{Q}{\rho}= -\frac{8 \pi Q}{M}\phi\left(\vec{x}\right),  \label{rn.gfa}
\end{equation}

Retaining terms up to leading order in $\phi$ and $1/l^2$, we then find that the metric components in \ref{rn.metb} take the form


\begin{align}
g_{00} = -\left(1 + 2 \phi+\frac{\rho^2}{l^2}\right) \,, \qquad g_{0i} = 0 \,, \qquad g_{ij} = \delta_{ij} \left(1 - 2\phi+\frac{\rho^2}{2l^2}\right)\,
\label{rn.meta}.
\end{align}

Note that the spacetime metric in \ref{rn.meta} provides the leading AdS correction about an asymptotically flat spacetime and in isotropic coordinates it just behaves like the AdS-Schwarzschild metric with a gauge potential. The metric is however equivalent to the metric in \ref{metric} up to leading order in $\phi$ and $1/l^2$. In particular, the timelike boundary of the full AdS spacetime is not part of the spacetime we have considered. In the remaining sections, we shall investigate the scattering of a charged probe particle in this background given by the metric in \ref{rn.meta} and gauge field in \ref{rn.gfa}.

\subsection{Perturbations of Einstein-Maxwell equations}

We now linearly perturb the spacetime by introducing a point probe particle with mass `$m$' and charge `$q$' moving along a worldline trajectory $r(\sigma)$, \cite{DeWitt:1960fc, Peters:1966, Peters:1970mx, Kovacs:1977uw, Poisson:2011nh} whose action is 
\begin{equation}
S_{P} = -m \int d\sigma \sqrt{-g_{\mu\nu}\frac{dr^{\mu}}{d \sigma}\frac{dr^{\nu}}{d \sigma}} + \frac{q}{4\pi} \int d\sigma A_{\mu}\frac{dr^{\mu}}{d \sigma}\,.
\label{act.pp}
\end{equation}
where $\frac{dr^{\mu}}{d \sigma}=u^{\mu}$ is the tangent to the worldline of the probe and the metric is evaluated at $r$. Variation of \ref{act.pp} gives the following stress tensor $T^{\mu \nu}_{(P)}$ {\footnote{We have chosen $\sigma$ to be proper time as measured in this spacetime and therefore $u^{\mu}$ satisfies the relation $g_{\mu\nu}u^{\mu}u^{\nu}=-1$}} and current $J^{\mu}_{(P)}$
\begin{align}
T^{\mu \nu}_{(P)} &=\frac{2}{\sqrt{-g}}\frac{\delta S_P}{\delta g_{\mu\nu}}=m\int \delta(x,r(\sigma))\frac{dr^{\mu}}{d \sigma}\frac{dr^{\nu}}{d \sigma}\,d \sigma \,, \notag \\
J_{(P)}^{\mu}&= \frac{1}{\sqrt{-g}}\frac{\delta S_P}{\delta A_{\mu}}=\frac{q}{4\pi} \int \delta(x,r(\sigma))\frac{dr^{\mu}}{d\sigma} \, d \sigma \,,
\end{align}
where $\delta(x,r(\sigma))$ is the covariant delta function. It is related to the flat spacetime delta function $\delta^{4}\left(x-r(\sigma)\right)$ via
\begin{equation}
\delta(x,r(\sigma)) \sqrt{-g} = \delta^{4}\left(x-r(\sigma)\right) = \delta(t - r^0(\sigma)) \delta^{(3)}(\vec{x} - \vec{r}(\sigma))\,,
\end{equation} and normalized as
\begin{equation}
\int \sqrt{-g}~\delta(x,r(\sigma)) d\sigma =1 \,.
\end{equation}
The stress-energy tensor and current of the point particle induces a perturbation of the background metric and gauge potential 
\begin{align}
g_{\mu\nu} &\to g_{\mu\nu}+ \delta g_{\mu\nu} = g_{\mu\nu} + 2 h_{\mu \nu}\,, \notag\\
A_{\mu} &\to A_{\mu}+ \delta A_{\mu} = A_{\mu} + a_{\mu}\,.
\end{align}
The variations of \ref{emm} and \ref{max} yield
\begin{align}
\delta \tilde{G}_{\mu\nu}  - \delta T^{h}_{\mu\nu} - \delta T^{a}_{\mu\nu} &=  T_{\mu\nu}^{(P)}\,,\label{eipert}\\
\delta(F^{\mu\nu};_{\nu})&=  4\pi J^{\mu}_{(P)} \,,\label{pert}
\end{align}
where 
\begin{equation}
\tilde{G}_{\mu\nu}=R_{\mu\nu}-\frac{1}{2}R g_{\mu\nu}+\Lambda g_{\mu\nu}\,.
\end{equation}
In \ref{eipert} we have split the total perturbation of the stress-energy tensor into two components, one part $\delta T^{h}_{\mu\nu}$ is due to the perturbation of the metric and another part $\delta T^{a}_{\mu\nu}$ is due to the perturbation of the gauge potential.

On simplifying $\delta \tilde{G}_{\mu\nu} \,, \delta T^{h}_{\mu \nu}\,, \delta T^{a}_{\mu\nu}$ and $\delta(F^{\mu\nu};_{\nu})$, we find
\begin{align}
\delta \tilde{G}_{\mu\nu}&= -e_{\mu\nu;\alpha}{}^{\alpha} + e_{\mu\alpha;}{}^{\alpha}{}_{\nu} + e_{\nu\alpha;\phantom{\alpha}\mu}^{\phantom{\nu\alpha;}\alpha}+ \left(R_{\nu}{}^{\delta}e_{\delta\mu}+R_{\mu}{}^{\delta}e_{\delta\nu}\right)+2R^{\alpha}{}_{\nu\mu}{}^{\delta}e_{\delta\alpha}  \notag\\& \quad -  g_{\mu\nu} e_{\alpha\beta;}{}^{\alpha\beta}+ g_{\mu\nu} R^{\alpha\beta} e_{\alpha\beta} - R e_{\mu\nu}+2\Lambda e_{\mu\nu}-\Lambda g_{\mu\nu}e \,, \notag\\
\delta T^{h}_{\mu\nu} &= e T_{\mu \nu}^{EM} -\frac{1}{8 \pi}e_{\mu\nu} F_{\alpha \beta} F^{\alpha \beta} - \frac{1}{2 \pi}g^{\alpha \epsilon}g^{\beta \delta}e_{\epsilon\delta}\left( F_{\alpha \mu} F_{\beta \nu} - \frac{1}{4} g_{\mu\nu} F_{\alpha \gamma} F_{\beta}{}^{\gamma}\right)\,,\notag\\
\delta T^{a}_{\mu\nu} &= \frac{1}{4 \pi}g^{\alpha \beta}\left( f_{\alpha \mu} F_{\beta \nu} + f_{\alpha \nu} F_{\beta \mu} - \frac{1}{2}g_{\mu\nu}g^{\gamma \delta} f_{\alpha \gamma} F_{\beta \delta}\right)\,,\notag\\
\delta(F^{\mu\nu};_{\nu}) & = - g^{\alpha \rho} g^{\mu \nu}\left[2 g^{\beta \sigma} \left(e_{\rho \sigma}F_{\nu \beta;\alpha}+e_{\sigma\nu;\alpha}F_{\beta\rho}\right) - e_{,\rho}F_{\nu\alpha} - f_{\nu \rho;\alpha} - 2 F_{\alpha \nu} e_{\rho \beta;}{}^{\beta} \right] \,,
\label{del.exp}
\end{align}
where we have denoted the perturbed electromagnetic field strength tensor by 
\begin{equation}
f_{\mu\nu}= a_{\nu,\mu} - a_{\mu,\nu} \,,
\end{equation}
and have introduced the trace-reversed metric perturbations $e_{\mu \nu}$ defined by
\begin{equation} 
e_{\mu\nu}=h_{\mu\nu}-\frac{1}{2}h g_{\mu\nu} \,; \qquad  h = g^{\mu \nu} h_{\mu \nu} = - e = - g^{\mu \nu} e_{\mu \nu}\,.
\end{equation}

Substituting the first three expressions of \ref{del.exp} in \ref{eipert}, we find the following expression for the perturbed Einstein equation 
\begin{align}
-T^{(P)}_{\mu \nu} &= e_{\mu\nu;\alpha}{}^{\alpha} - e_{\mu\alpha;}{}^{\alpha}{}_{\nu} - e_{\nu\alpha;\phantom{\alpha}\mu}^{\phantom{\nu\alpha;}\alpha} - \left(R_{\nu}{}^{\delta}e_{\delta\mu}+R_{\mu}{}^{\delta}e_{\delta\nu}\right) - 2R^{\alpha}{}_{\nu\mu}{}^{\delta}e_{\delta\alpha} + R e_{\mu\nu}-2\Lambda e_{\mu\nu}\notag\\& \quad +\Lambda g_{\mu\nu}e - g_{\mu\nu} R^{\alpha\beta} e_{\alpha\beta} + g_{\mu\nu} e_{\alpha\beta;}{}^{\alpha\beta} + \frac{1}{4\pi} g^{\alpha \beta}\left( f_{\alpha \mu} F_{\beta \nu} + f_{\alpha \nu} F_{\beta \mu} - \frac{1}{2}g_{\mu\nu}g^{\gamma \delta} f_{\alpha \gamma} F_{\beta \delta}\right)\notag\\ & \qquad +  e T_{\mu \nu}^{EM} - \frac{1}{8\pi}e_{\mu\nu} F_{\alpha \beta} F^{\alpha \beta} - \frac{1}{2\pi} g^{\alpha \epsilon}g^{\beta \delta}e_{\epsilon\delta}\left( F_{\alpha \mu} F_{\beta \nu} - \frac{1}{4} g_{\mu\nu} F_{\alpha \gamma} F_{\beta}{}^{\gamma}\right)\,.
\label{ein.pertha}
\end{align}
Similarly plugging the expression of $\delta(F^{\mu\nu};_{\nu})$ from \ref{del.exp} in \ref{pert} gives the perturbed Maxwell equation
\begin{equation}
-4\pi g^{\mu \nu} J_{\nu}^{(P)}= g^{\alpha \rho} g^{\mu \nu}\left[2 g^{\beta \sigma} \left(e_{\rho \sigma}F_{\nu \beta;\alpha}+e_{\sigma\nu;\alpha}F_{\beta\rho}\right) - e_{,\rho}F_{\nu\alpha} - f_{\nu \rho;\alpha} - 2 F_{\alpha \nu} e_{\rho \beta;}{}^{\beta} \right].
\label{max.pertha}
\end{equation}
We will now express \ref{ein.pertha} and \ref{max.pertha} about the background with the metric \ref{rn.meta} and gauge potential \ref{rn.gfa}. We will also rewrite parts of the equations in terms of the following quantities, 
\begin{equation}
k_{\mu} = e_{\mu\nu ;}{}^{\nu} \,, \qquad b = -a_{0,0} + a_{i,i}\,.
\label{gauge}
\end{equation}
The radiative components of gravitational and electromagnetic perturbations are spatial in isotropic coordinates. The spatial components of \ref{ein.pertha} are
\begin{align}
- T^{(P)}_{i j} &= \Box\left(\left(1+2\phi - \frac{\rho^2}{2 l^2}\right) e_{ij}\right) - k_{i,j} - k_{j,i} - 2(\left(\phi_{,k} - \frac{\rho^2_{,k}}{4 l^2}\right) e_{ki})_{,j} - 2(\left(\phi_{,k} - \frac{\rho^2_{,k}}{4 l^2}\right)e_{kj})_{,i}   \notag\\
& \quad - \left(k_{0,0} - k_{l,l}\right) \delta_{ij} +4 \left[ \phi e_{ij,00} + \phi_{,i} e_{0j,0}+ \phi_{,j} e_{i0,0} +\frac{1}{2}\left(\phi_{,ij} - \frac{1}{2}\phi_{,kk}\delta_{ij}\right) \left(e_{00}+ e_{ll}\right)\right]\notag\\ 
& \qquad + 2 \delta_{ij}\left(\phi_{,kl} e_{kl} - \phi_{,k}k_{k}+ 2 \phi k_{0,0}\right) + \frac{\delta_{ij}}{4l^2}\left(\rho^2_{,kl} e_{kl} + \rho^2_{,k}k_{k}+ 2 \rho^2 k_{0,0}\right)  \notag\\
&\qquad \; +\frac{1}{2l^2}\left[\rho^2 e_{ij,00}+\rho^2_{,i}e_{j0,0}+\rho^2_{,j}e_{i0,0}+\rho_{,ij}^2(2e_{00}-e_{ll})-\frac{1}{2}\delta_{ij}\rho^2_{,kk}e_{ll}\right.\notag\\
&\left. \qquad \quad +\frac{3}{2}(\rho^2_{,ki}e_{kj}+\rho^2_{,kj}e_{ki})-\rho^2_{,kk}e_{ij}+\frac{3}{2}\rho_{,k}^2e_{ij,k}\right] - \frac{2 Q}{M}\left(f_{0i}\phi_{,j}+f_{0j}\phi_{,i} - \delta_{ij}f_{0l}\phi_{,l}\right),
\label{ein.pertf}
\end{align}
where $\Box=-\partial_0^2+\partial_i^2$. The spatial component of perturbed Maxwell equation \ref{max.pertha} in the AdS-Reissner-Nordstr\"om background gives
\begin{align}
&-4\pi J^{i}_{(P)}= -4\pi \left(1 + 2 \phi - \frac{\rho^2}{2l^2}\right)J_{i}^{(P)}\notag\\
&\, = \Box a_i - b_{,i} + \frac{3 \rho^2}{2 l^2} \left(a_{i,00} - a_{0,0i}\right)+ \left(4 \phi - \frac{\rho^2}{l^2}\right) \left(a_{i,kk} - a_{k,ki}\right) - 2\left(\phi_{,k} + \frac{\rho^2_{,k}}{8 l^2}\right) f_{ik} \notag\\
&\qquad + \frac{16 \pi Q}{M}\left(e_{j 0} \phi_{,ij} + \left( e_{ij,0} - e_{i0,j}\right)\phi_{,j} + \frac{1}{2}\left(e_{0 0,0} - e_{kk,0}\right)\phi_{,i} + \phi_{,i}k_0\right)\label{tum}.
\end{align}
We now need to implement gauge choices. To this end, we adopt the following choice for $k_{\mu}$ and $b$ to simplify equations \ref{ein.pertf} and \ref{tum}
\begin{align}
k_{\mu}&=-2\left(\phi_{,k}-\frac{\rho_{,k}^2}{4l^2}\right)e_{k\mu}+ \frac{2Q}{M} a_0 \phi_{,\mu}\,,
\label{g.gauge}\\b &= - 2\left(\phi_{,k} + \frac{\rho^2_{,k}}{8 l^2}\right) a_k - 4\left(\phi + \frac{\rho^2}{8 l^2}\right) a_{k,k}.
\label{a.gauge}
\end{align}
Using \ref{g.gauge} one can simplify \ref{ein.pertf} to
\begin{align}
& \Box\left(\left(1+2\phi - \frac{\rho^2}{2 l^2}\right) e_{ij}\right) +4 \left[ \phi e_{ij,00} + \phi_{,i} e_{0j,0}+ \phi_{,j} e_{i0,0} +\frac{1}{2}\left(\phi_{,ij} - \frac{1}{2}\phi_{,kk}\delta_{ij}\right) \left(e_{00}+ e_{ll}\right)\right]\notag\\ 
& +\frac{1}{2l^2}\left[\rho^2 e_{ij,00}+\rho^2_{,i}e_{j0,0}+\rho^2_{,j}e_{i0,0}+\rho_{,ij}^2(2e_{00}-e_{ll})-\frac{1}{2}\delta_{ij}\rho^2_{,kk}e_{ll} +\frac{3}{2}(\rho^2_{,ki}e_{kj}+\rho^2_{,kj}e_{ki}) +\frac{3}{2}\delta_{ij}\rho^2_{,kl}e_{kl}\right. \notag\\
&\left. -\rho^2_{,kk}e_{ij}+\frac{3}{2}\rho_{,k}^2e_{ij,k}\right] - \frac{2 Q}{M}\left[a_{i,0}\phi_{,j}+a_{j,0}\phi_{,i} - \delta_{ij}a_{l,0}\phi_{,l} + 2 a_0\left(\phi_{,ij} - \frac{1}{2}\delta_{ij}\phi_{,kk}\right)\right]=-  T^{(P)}_{i j} .
\label{ein.perf2}
\end{align}
Similarly for the perturbed Maxwell equation \ref{tum} we use both \ref{g.gauge} and \ref{a.gauge} to get
\begin{align}
&-4\pi \left(1 - \phi + \frac{5\rho^2}{8l^2}\right)J_{i}^{(P)} = \Box\left(\left(1 + \phi + \frac{\rho^2}{8l^2}\right)a_i\right) \notag\\
&\quad + 4\left[\left(\phi +\frac{\rho^2}{8 l^2}\right) a_{i,00}+ \left(\phi_{,i} + \frac{\rho^2_{,i}}{8l^2}\right)a_{0,0}\right]
+ 2\left(\phi_{,ki} + \frac{\rho^2_{,ki}}{8l^2}\right)a_k  - \left(\phi_{,kk} + \frac{\rho_{,kk}^2}{8l^2}\right)a_i\notag\\&  \qquad + \frac{16 \pi Q}{M} \left[e_{j 0} \phi_{,ij}+\left( e_{ij,0} - e_{i0,j}\right)\phi_{,j} + \frac{1}{2}\left(e_{0 0,0} - e_{kk,0}\right)\phi_{,i}\right]\,.
\label{max.fin}
\end{align}
Note that for both the equations \ref{ein.perf2} and \ref{max.fin} we keep the terms up to leading order in $\phi$ and $1/l^2$. In the next section we will solve \ref{ein.perf2} and \ref{max.fin} for $e_{ij}$ and $a_i$ in frequency space using the worldline formalism. The result will involve a Green's function for the $1/l^2$ correction which was previously derived in  \cite{Banerjee:2020dww}.

\section{Solutions of the field equations}\label{sec3}
To solve \ref{ein.perf2} and \ref{max.fin} we first briefly review the solution for a perturbed scalar field equation. For an arbitrary source $f(\sigma)$, the solution of the following scalar box equation in a curved spacetime
\begin{equation}
\psi_{;\alpha}{}^{\alpha}(x) = -\int \delta(x,r(\sigma)) f(\sigma) d\sigma\,.
\label{scalar.cov}
\end{equation}
can be written as  
\begin{align}
\psi^{(1)}(x) &= \psi^{(0)}(x) + \delta \psi^{(0)}(x) \notag\\
&= \frac{1}{4\pi}\int \limits_{-\infty}^{\sigma_0} \delta\left(-\Omega\left(x,r(\sigma)\right)\right) f(\sigma) d\sigma\notag\\
& \; + \frac{1}{16 \pi^2} \int \sqrt{-g(y)} \delta\left(-\Omega\left(x,y\right)\right) d^4y  \int \limits_{-\infty}^{\sigma_0} \delta'\left(-\Omega\left(y,r(\sigma)\right)\right) F\left(y,r(\sigma)\right) f(\sigma) d\sigma\,.
\label{psi1.sol}
\end{align}
where $\Omega\left(x,r(\sigma)\right)$ is the Synge world function and $F\left(x,r(\sigma)\right)$ is the Ricci tensor dependent term which arises from derivatives of the world function {\footnote{For a more detailed discussion on the world function in the context of our derivation we refer the reader to \cite{Banerjee:2020dww}.}}
\begin{align}
\Omega(x,r)=\frac{1}{2}(u_1-u_0)\int_{u_0}^{u_1}~g_{\alpha\beta}U^{\alpha}U^{\beta}~du\label{wf},\\
F\left(x,r\right) = \frac{1}{u_1 - u_0} \int \limits_{u_0}^{u_1} \left(u-u_0\right)^2 R_{\mu \nu} U^{\mu} U^{\nu} du\,\label{F}.
\end{align}
In \ref{wf} and \ref{F} we assume that the observer ($x$) and probe particle source ($r$) are joined by a unique geodesic $\xi^{\alpha}$ with affine parameter `u' and $U^{\alpha}=\frac{d\xi^{\alpha}}{du}$ is the tangent vector to the geodesic. Gravitational and electromagnetic radiation follow this path from the source to the observer. $R_{\mu \nu}$ in \ref{F} gets the contribution from the black hole, which in our scattering approximation can be treated as a point particle with mass $M$ and charge $Q$. The integration limit is chosen up to $\sigma_0$ instead of infinity $(\infty)$. This ensures that $r^{\mu}(\sigma_0)$ lies outside the light cone centred at $x^{\mu}$ and the contribution to the scalar pertubation $\psi$ only comes from the retarded part of the Green's function.

Expanding \ref{scalar.cov} in terms of the d'Alembertian operator we get
\begin{align}
 \Box \psi^{(1)} + 4 \phi \partial_t^2 \psi^{(0)}-\frac{\rho^2}{4l^2}\partial_t^2 \psi^{(0)}&+\frac{3}{4l^2}\rho^2_k \partial_k \psi^{(0)}+\frac{3\rho^2}{4l^2}\partial_k^2 \psi^{(0)}\notag\\&= - \int \delta^{4}\left(x - z(\sigma)\right) f(\sigma) d\sigma + \mathcal{O}\left(R^2\right).
\label{scalar.fe}
\end{align}
We can solve \ref{scalar.fe} by substituting \ref{psi1.sol} and performing a Fourier transformation
\begin{equation}
\tilde{\psi}^{(1)} \left(\omega, \vec{x}\right) = \int dt e^{i\omega t} \psi^{(1)} \left(t, \vec{x}\right).
\end{equation}

The transformed field $\tilde{\psi}$ can be perturbatively solved about flat spacetime. The solution of $\tilde{\psi}$ that are leading order in $\phi$ and $1/l^2$ provide tail contributions to the flat spacetime Green's function, which arises due to the black hole potential and AdS potential. Denoting these tail terms as $G_M$ and $G_l$, they have the solutions~\cite{Fernandes:2020tsq, Banerjee:2020dww}
\begin{align}
G_M\left(\omega,\vec{x},\vec{r}\right) &= -\frac{i M}{16 \pi \omega} \left(\frac{e^{i \omega R_0} \Gamma\left(\vec{x},\vec{r}\right)}{R_0} - \int \limits_{0}^{\infty} dv \frac{e^{i \omega\left(v + \vert\vec{z}\vert + \rho(v)\right)}}{\left(v+\vert \vec{z}\vert\right)\rho(v)} \right) \,, \label{tail.m}\\G_l\left(\omega,\vec{x},\vec{r}\right) &= -~\frac{4i}{\omega}~ e^{i \omega R_0} \left(\vec{x}.\vec{r}\right) \,, \label{tail.l}
\end{align}
where $G_M$ and $G_l$ correspond to the contribution due to the black hole mass and AdS radius respectively, $\vec{R}_0 = \vec{x} - \vec{r}(\sigma)$ with magnitude $R_0 = \vert \vec{R}_0\vert$, and the following expressions for $\Gamma\left(\vec{x},\vec{r}\right)$ and  $\rho(v)$

\begin{equation}
 \Gamma\left(\vec{x},\vec{r}\right) = \text{ln} \left(\frac{\vert \vec{x}\vert R_0 + \vec{x}.\vec{R}_0 }{\vert\vec{r}\vert R_0 + \vec{r}.\vec{R}_0}\right) \,, \qquad \rho(v) = \sqrt{x^2 + v^2 + \frac{2 v \vec{x}.\vec{r}}{\vert \vec{r}\vert}}\,.
\label{gamr.def}
\end{equation}

The equations for the perturbed fields $e_{ij}$ and $a_i$ have additional contributions from the background apart from those in \ref{scalar.fe} due to their respective tensor and vector nature. These additional contributions are frequency space solutions in terms of the derivatives of \ref{tail.m} and \ref{tail.l}
\begin{equation}
-\nabla_i \tilde{\Box} G_M \left(\omega,\vec{x},\vec{r}\right) = \phi_{,i}\frac{e^{i\omega R_0}}{R_0} \,, \quad -\nabla_i \nabla_k \tilde{\Box} G_M \left(\omega,\vec{x},\vec{r}\right) = \phi_{,ik} \frac{e^{i\omega R_0}}{R_0}\,,\label{Dl} 
\end{equation}
and
\begin{equation}
-\nabla_i \tilde{\Box} G_l \left(\omega,\vec{x},\vec{r}\right) = \rho^2_{,i}\frac{e^{i\omega R_0}}{R_0} \,, \quad -\nabla_i \nabla_k \tilde{\Box} G_l \left(\omega,\vec{x},\vec{r}\right) = \rho^2_{,ik} \frac{e^{i\omega R_0}}{R_0} \,,\label{DG}
\end{equation}
where $\nabla_i = \frac{\partial}{\partial x^i} + \frac{\partial}{\partial r^i} $ is an operator that acts on the two spatial arguments in the Green's functions and $\tilde{\Box}=(\omega^2+\partial_i^2)$.

We can now derive the frequency space solutions $\tilde{e}_{ij}$ and $\tilde{a}_i$. This requires substituting $e_{\mu\nu}$ and $a_{\mu}$ in terms of their Fourier transformed fields $\tilde{e}_{\mu\nu}\left(\omega\,,\vec{x}\right)$ and $\tilde{a}_{\mu}\left(\omega\,,\vec{x}\right)$ in \ref{ein.perf2} and \ref{max.fin}. In the case of \ref{ein.perf2}, we find
\begin{align}
 -T^{(P)}_{i j} &= \int d\omega\, e^{-i \omega t}\, \widetilde{\Box}\left(\left(1+2\phi-\frac{\rho^2}{2l^2}\right) \tilde{e}_{ij}\right) \notag\\
&- \int d\omega\, e^{-i \omega t}\, 4 \left[ \omega^2 \phi \tilde{e}_{ij} + i \omega \left(\phi_{,i} \tilde{e}_{j0} + \phi_{,j} \tilde{e}_{i0}\right) -\frac{1}{2}\left(\phi_{,ij} - \frac{1}{2}\phi_{,kk}\delta_{ij}\right) \left(\tilde{e}_{00}+ \tilde{e}_{ll}\right)\right]\notag\\&-\frac{1}{2l^2}\left[\omega^2\rho^2 \tilde{e}_{ij}+i\omega\rho^2_{,i}\tilde{e}_{j0}+i\omega\rho^2_{,j}\tilde{e}_{i0}-\rho_{,ij}^2(2\tilde{e}_{00}-\tilde{e}_{ll})-\frac{3}{2}\delta_{ij}\rho^2_{,kl}\tilde{e}_{kl}+\frac{1}{2}\delta_{ij}\rho^2_{,kk}\tilde{e}_{ll}\right.\notag\\&\left.\qquad-\frac{3}{2}(\rho^2_{,ki}\tilde{e}_{kj}+\rho^2_{,kj}\tilde{e}_{ki})-\rho^2_{,kk}\tilde{e}_{ij}-\frac{3}{2}\rho_{,k}^2\tilde{e}_{ij,k}\right] \notag\\
&\qquad + \frac{2 Q}{M}\int d\omega\, e^{-i \omega t} \left[i \omega \left(\tilde{a}_{i}\phi_{,j} + \tilde{a}_{j}\phi_{,i} - \delta_{ij} \tilde{a}_{l}\phi_{,l}\right) - 2 \tilde{a}_0\left(\phi_{,ij} - \frac{1}{2}\delta_{ij}\phi_{,kk}\right)\right]\,,
\label{ft.grav}
\end{align}
where
\begin{align}
T^{(P)}_{i j}=m \int \delta_{ki}\delta_{lj}~\frac{\delta^4(x-r(\sigma))}{1+2\phi(\vec{r})+\frac{r^2}{4l^2}}\frac{dr^{k}}{d\sigma}\frac{dr^{l}}{d\sigma}\,d\sigma \,.
\end{align}
Similar steps for the perturbed Maxwell equation \ref{max.fin} gives
\begin{align}
&-\left(1 - \phi +\frac{5\rho^2}{8l^2}\right) 4 \pi J^{(P)}_i = -q\int \delta_{ki}\frac{\delta^4(x - r(\sigma))}{1 + \phi\left(\vec{r}\right)+\frac{r^2}{8l^2}}\frac{dr^k}{d \sigma}\,d \sigma\notag\\&  \hspace{10 em}=\int d\omega\, e^{-i \omega t}\, \widetilde{\Box}\left(\left(1+\phi+\frac{\rho^2}{8l^2}\right)\tilde{a}_i\right)\notag\\& \hspace{5 em}- \int d\omega\, e^{-i \omega t} \left[ 4 \left(\omega^2 \tilde{a}_{i}\phi + i \omega  \tilde{a}_{0}\phi_{,i}\right) + \tilde{a}_i\phi_{,kk} - 2\tilde{a}_k\phi_{,ik}\right]\notag\\&\hspace{5 em} +\int d\omega\, e^{-i \omega t} \frac{16 \pi Q}{M}\left(\tilde{e}_{0j}\phi_{,ij}-\left( i\omega \tilde{e}_{ij} + \tilde{e}_{i0,j}\right)\phi_{,j} - \frac{1}{2}\left(i\omega \tilde{e}_{0 0} - i\omega\tilde{e}_{kk}\right)\phi_{,i}\right)\notag\\&\hspace{10 em}- \int d\omega\, e^{-i \omega t} \left[ 4 \left(\omega^2 \tilde{a}_{i}\frac{\rho^2}{8l^2} + i \omega  \tilde{a}_{0}\frac{\rho^2_{,i}}{8l^2}\right) + \tilde{a}_i\frac{\rho^2_{,kk}}{8l^2} - \tilde{a}_k\frac{\rho^2_{,ik}}{4l^2} \right]\,.
\label{ft.em}
\end{align}
We get the solution for $\tilde{e}_{ij}$ from the scalar perturbation solution comparing $\tilde{\psi}^{(0)}_0(\omega\,,\vec{x})$ with $ \left(1 + 2 \phi \left(\vec{x}\right) - \frac{x^2}{2l^2}\right) \tilde{e}_{ij}^{(0 )}\left(\omega\,,\vec{x}\right)$ and replacing $f(\sigma)$ with $2 m\delta_{ki}\delta_{lj}
 \left(1 - 2\phi\left(\vec{r}\right) - \frac{r^2}{4l^2}\right)\frac{dr^k}{d \sigma}\frac{dr^l}{d \sigma}$. The zeroth order solution of $\tilde{e}_{ij}(\omega\,,\vec{x})$ in frequency space is
\begin{equation}
\tilde{e}^{(0)}_{ij}(\omega\,,\vec{x}) =m \int \frac{e^{i \omega\left(r^0 + R_0\right)}}{4 \pi R_0} v_i v_j \, \frac{dr^0}{d\sigma}dr^0 \,,
\label{lgrav.sol}
\end{equation}
where we have denoted $\frac{dr^k}{d r^0}$ as $v^k$.

Likewise, comparing $\tilde{\psi}^{(0)}_0(\omega\,,\vec{x})$ to $ \left(1 + \phi \left(\vec{x}\right)+\frac{x^2}{8l^2}\right) \tilde{a}_{i}\left(\omega\,,\vec{x}\right)$ and replacing $f(\sigma)$ with \\ $q \delta_{ki} \left(1-\phi\left(\vec{r}\right)-\frac{r^2}{8l^2}\right)\frac{dr^k}{d \sigma}$, we find the following zeroth order solution $\tilde{a}^{(0)}_{i}(\omega\,,\vec{x})$ in Fourier space
\begin{equation}
\tilde{a}^{(0)}_{i}(\omega\,,\vec{x}) = q \int \frac{e^{i \omega\left(r^0 + R_0\right)}}{4 \pi R_0} v_i \,dr^0 \,.
\label{lem.sol}
\end{equation}
We can further compute the other components of gravitational and electromagnetic perturbations from $\tilde{e}_{ij}$ and $\tilde{a}_i$, as they are related among themselves by  the gauge fixing condition. It follows from our choice in \ref{g.gauge} and \ref{a.gauge} that \ref{gauge} on flat spacetime simplifies to
\begin{align}
e_{ij,j} - e_{i0,0} &= 0 \,, \qquad
e_{0i,i} - e_{00,0} =0 \,,\label{ef}\\
& a_{0,0} - a_{i,i} = 0.
\label{lor.eq}
\end{align}
These are simply the de Donder and Lorenz gauges in flat spacetime. By Fourier transforming \ref{ef} and using \ref{lgrav.sol}, we can now derive the following zeroth order solutions of $\tilde{e}_{i0}$ and $\tilde{e}_{00}$
\begin{align}
\tilde{e}^{(0)}_{i0}(\omega\,,\vec{x}) &= -m \int \frac{e^{i \omega\left(r^0 + R_0\right)}}{4 \pi R_0} v_i  \, \frac{dr^0}{d\sigma}dr^0 + \mathcal{O}(\phi)\,,\notag\\
\tilde{e}^{(0)}_{00}(\omega\,,\vec{x}) &= m \int \frac{e^{i \omega\left(r^0 + R_0\right)}}{4 \pi R_0}  \, \frac{dr^0}{d\sigma}dr^0 + \mathcal{O}(\phi)\,.
\label{low.e}
\end{align}
Using the Fourier transform of \ref{lor.eq}, we can similarly use \ref{lem.sol} to determine the solution for electromagnetic perturbation $\tilde{a}^{(0)}_{0}$
\begin{align}
\tilde{a}^{(0)}_{0}(\omega\,,\vec{x}) &= -q \int \frac{e^{i \omega\left(r^0 + R_0\right)}}{4 \pi R_0} dr^0 + \mathcal{O}(\phi)\,.
\label{low.a}
\end{align}
Hence the gauge conditions determine all the lowest order expressions. To find the complete solution, we first substitute all zeroth order solutions \ref{lgrav.sol}, \ref{low.e}, \ref{lem.sol} and \ref{low.a} in all terms that are coefficients of $\phi$ and $1/l^2$ in \ref{ft.grav}. We then use the expressions in \ref{DG} and \ref{Dl} to determine the  following solution for $\tilde{e}_{ij}\left(\omega,\vec{x}\right)$

\begin{align}
\tilde{e}_{ij}\left(\omega,\vec{x}\right) &= \frac{m}{1 + 2 \phi(\vec{x})-\frac{x^2}{2l^2}} \int dr^0 \,\frac{dr^0}{d\sigma}\,\frac{e^{i \omega\left(r^0 + R_0\right)}}{4 \pi R_0} \frac{v_i v_j}{1+ 2 \phi(\vec{r})+\frac{r^2}{4l^2}}  \notag\\
&\quad - \int dr^0\, e^{i \omega r^0} \int d^3\vec{r}\,'\, \delta^{(3)} \left(\vec{r}\,' - \vec{r}\left(r^0\right)\right) \Bigg\{\frac{dr^0}{d\sigma} \frac{m}{\pi} \left[\omega^2 v^i v^j - i \omega \left(v^i\nabla_j + v^j\nabla_i\right) \phantom{\left(\frac{1}{2}\right)} \right. \notag\\
&\left. \qquad - \frac{\left(1+\vec{v}^2\right)}{2}\left(\nabla_i \nabla_j - \frac{1}{2}\delta_{ij}\nabla_k\nabla_k\right)\right] + \frac{q Q}{2 \pi M} \left[i\omega \left( v^i \nabla_j + v^j\nabla_j - \delta_{ij} v^k \nabla_k\right) \phantom{\left(\frac{1}{2}\right)}\right. \notag\\
& \left. \qquad \qquad \quad \, + 2 \left(\nabla_i \nabla_j - \frac{1}{2}\delta_{ij}\nabla_k \nabla_k\right)\right]\Bigg\} G_M\left(\omega,\vec{x},\vec{r}\,'\right)\notag\\
&\quad - \int dr^0\, e^{i \omega r^0} \int d^3\vec{r}\,'\, \delta^{(3)} \left(\vec{r}\,' - \vec{r}\left(r^0\right)\right) \Bigg\{\frac{dr^0}{d\sigma}\frac{m}{8\pi l^2} \left[\omega^2 v_i v_j - i \omega \left(v_i\nabla_j + v_j\nabla_i\right)\right.\notag\\&\left.-\left(2-\vec{v}^2\right)\nabla_i \nabla_j\right.- \frac{3}{2}\delta_{ij}v_k v_m\nabla_k\nabla_m- \frac{1}{2}\delta_{ij}v^2\nabla_k\nabla_k\notag\\&\left. - \frac{3}{2}\left(v_k v_j\nabla_k\nabla_i+v_k v_i\nabla_k\nabla_j\right)+v_iv_j\nabla_k \nabla_k+\frac{3}{8}i\omega(v_i\nabla_j+v_j\nabla_i)\right]\Bigg\} G_l\left(\omega,\vec{x},\vec{r}~'\right)\,.
\label{hija.fin}
\end{align}
Carrying out similar steps for $\tilde{a}_{i}\left(\omega,\vec{x}\right)$, from \ref{ft.em} we get
\begin{align}
&\tilde{a}_{i}\left(\omega,\vec{x}\right) = \frac{q}{1 + \phi(\vec{x})+\frac{x^2}{8l^2}} \int dr^0\,\frac{e^{i \omega\left(r^0 + R_0\right)}}{4 \pi R_0} \frac{v_i }{1+ \phi(\vec{r})+\frac{r^2}{8l^2}} \notag\\
&\quad - \int dr^0\, e^{i \omega r^0} \int d^3\vec{r}\,'\, \delta^{(3)} \left(\vec{r}\,' - \vec{r}\left(r^0\right)\right) \Bigg\{\frac{q}{\pi} \left[ \omega^2 v_i - i \omega \nabla_i  + \frac{1}{4} v_i\nabla_k\nabla_k - \frac{1}{2} v^k \nabla_k \nabla_i \right]  \notag\\
& \qquad \qquad \qquad + \frac{dr^0}{d\sigma}\frac{4 Q m}{M}\left[ v^j\nabla_i\nabla_j + i \omega\left(v^i v^j \nabla_j + \frac{1}{6}\nabla_i - \frac{1}{2}\vec{v}^2 \nabla_i\right)\right]  \Bigg\} G_M\left(\omega,\vec{x},\vec{r}\,'\right)\notag\\
&- \frac{q}{4\pi}\frac{1}{l^2} \int dr^0\, e^{i \omega r^0} \int d^3\vec{r}\,'\, \delta^{(3)} \left(\vec{r}\,' - \vec{r}\left(r^0\right)\right) \Bigg\{ \frac{1}{2}\omega^2\ v_i -\frac{1}{2} i \omega \nabla_i+ \frac{1}{8} v_i\nabla_k\nabla_k - \frac{1}{4} v_k \nabla_k \nabla_i \Bigg\}\notag\\&\hspace{10 em} G_l\left(\omega,\vec{x},\vec{r}\,'\right)\,.
\label{aia.fin}
\end{align}
In the next section we explicitly carry out the soft expansion of $\tilde{a}_i$ following the prescription described in \cite{Fernandes:2020tsq}. We will see that the contribution to the soft photon factor due to the AdS radius will only come from the first line of \ref{aia.fin}. The terms in the last line of \ref{aia.fin} give finite contributions in considering the double scaling limit of a vanishing frequency and infinite AdS radius.

\section{Classical soft photon factor}\label{sec4}
Computing the quantum soft factor of a quantum scattering amplitude in asymptotically AdS spacetime is tricky. Asymptotic states cannot be defined in AdS as it has timelike boundary and particle geodesics are periodic. Therefore we choose to calculate the soft factor from a classical prescription. One can compute the soft factors for photons or gravitons in asymptotically flat spacetimes by considering the classical limit of single/multiple soft theorems arising from a quantum scattering process \cite{Laddha:2018myi}. The same factors can also be derived from the low frequency classical radiation produced in a classical scattering process. The classical scattering is subject to the condition that the total energy carried by the soft radiation is small compared to the energy carried by the scatterer. Finally in the classical limit, the soft factor is extracted from the power spectrum of the low frequency classical radiation. 
 
Considering that the observer is far away from the probe, i.e. $\vec{x}\gg\vec{r}(\sigma)$ and taking the frequency $\omega\rightarrow 0$ limit in 4 spacetime dimensions on asymptotically flat backgrounds, the classical radiative field for the photon can be written in terms of the soft factors as \cite{Laddha:2018rle}, \cite{Laddha:2018myi}
\begin{equation}
\epsilon^{\alpha}\, \tilde a_{\alpha}(\omega, \vec x) = {\cal{N}}'\,S_{\rm em}(\epsilon, k),  \, \label{sfa}
\end{equation}
where $\epsilon^{\alpha}$ is an arbitrary polarization vector of the photon, $S_{\rm em}$ is the soft photon factor and `$k$' denotes the momentum of soft photon. Similarly for the soft graviton factor one can write
\begin{align}
\epsilon^{\alpha\beta}\, \tilde e_{\alpha\beta}(\omega, \vec x) = {\cal{N}}'\,S_{\rm gr}(\epsilon, k) \, . \label{sfg}
\end{align}
In \ref{sfg}, $\epsilon^{\alpha\beta}$ is an arbitrary rank two polarization tensor of the graviton, $S_{\rm gr}$ is the soft graviton factor and `$k$' denotes the momentum of the soft graviton \cite{Laddha:2018myi}. For both \ref{sfa} and \ref{sfg}
\begin{align}
R \equiv |\vec x|, \quad {\cal{N}}' \equiv -\frac{i}{4\pi} {e^{i\omega R}\over R}, \quad k \equiv -\omega(1, \hat n), \quad \hat n={\vec x\over R}\label{soft}.
\end{align}
 The soft factors $S_{\rm em}$ and $S_{\rm gr}$ have a term proportional to $\omega^{-1}$ at leading order in frequency and another term proportional to $\ln\omega^{-1}$ at subleading order\footnote{A detailed description of how to derive $S_{\rm gr}$ is given in \cite{Laddha:2018rle, Laddha:2018myi, Laddha:2019yaj, Saha:2019tub}.}. Equations \ref{sfa} and \ref{sfg} can be considered as an alternate definition for the soft factor which can be easily computed from considering the soft limit of classical electromagnetic and gravitational radiation profiles. 
 
In the case of asymptotically flat backgrounds, to calculate the soft factor one needs to consider the large $|t|$ limit and a suitable parametrization of $\vec r(t)$, where $\vec{r}(t)$ is the position of the scattered probe particle at time $t$. As shown in \cite{Banerjee:2020dww}, for computing the soft factor up to $1/l^2$ in an asymptotically AdS spacetime, we can still consider the particle to follow an approximately straight line geodesic for large values of $t$. In particular due to long range interaction caused by black hole potential and AdS potential, the trajectory receives following contribution
\begin{equation}
\vec{r}(t) \sim c_1 \ln |t|-\frac{c_2}{l^2}t^2 \,.\label{tlr}
\end{equation} 
The second term in the RHS of \ref{tlr} will effectively contribute $1/(\omega^2l^2)$ in frequency space. This is a constant contribution in the limit that we will be interested in, which will be further elaborated in this section. As the contribution from the AdS potential to the soft factor is beyond the $\ln\omega^{-1}$ subleading order in frequency, we do not require its correction to the particle trajectory for computing the soft factor up to this order in $1/l^2$. 

We hence parametrize $\vec r(t)$ for asymptotic trajectories at large $|t|$ in four spacetime dimensions similar to the flat spacetime case as \cite{Laddha:2018myi}
\begin{equation}
\vec{r}(t)=\vec{\beta}_{\pm}t -C_{\pm}\,\vec{\beta}_{\pm}\,\ln\vert t\vert+\,\text{finite terms},\qquad\vec{v}=\vec{\beta}_{\pm}\left(1-\frac{C_{\pm}}{t}\right).\label{asymrv}
\end{equation}
where $C_{\pm}$ are constants and $t$ denotes the proper time. The $\ln\vert t\vert$ terms are the contributions from long range interaction forces which only exist in 4 spacetime dimensions. \ref{asymrv} can be written in the following covariant form
\begin{equation}
    r^{\mu}_{(a)}(t)=\eta_{(a)}\frac{1}{m_{(a)}}p_{(a)}^{\mu}t+c_{(a)}^{\mu}\ln|t|\,,
\end{equation}
where $\eta_{(a)}$ is positive (negative) for incoming (outgoing) particles and $m_{(a)}$ is the mass of the a-th particle. We will consider the proper time as negative for incoming particles and positive for outgoing particles. Using the parametrization \ref{asymrv}, we will retain terms up to $1/t$ as these are relevant in the soft expansion of $\tilde{e}_{ij}$ and $\tilde{a}_i$. Using suitable integrals given in \cite{Laddha:2018myi} one can easily find the soft factors following the relation \ref{sfa} and \ref{sfg}.

For asymptotic AdS spacetimes, soft limits cannot merely imply a vanishing frequency limit. Since there is no notion of null infinity in asymptotic AdS spacetimes and any massless radiation gets bounced off an infinite number of times from spatial infinity, the frequency of a radiation can never strictly go to zero. To define a ``soft limit" in this case one needs to consider a double scaling limit. In \cite{Banerjee:2020dww} the soft limit for AdS was defined as simultaneously taking the frequency of the radiation in AdS space to zero and the radius of AdS space to infinity i.e. $\omega\rightarrow 0$ and $l\rightarrow \infty$, keeping $\omega l$ fixed. Another interesting feature of the metric \ref{rn.meta} is that the effect of long range interactions are the same as in asymptotically flat spacetimes \cite{Banerjee:2020dww}. This happens because the perturbed fields already contain terms up to $1/l^2$ and we are interested in results up to this order only. As will be further explained in Sec.~\ref{sec6}, the large `$l$' limit of an asymptotically AdS spacetime defines an asymptotically flat spacetime patch near the center of the spacetime. Our $\frac{1}{l^2}$ correction results concern scattering processes at short times, with asymptotic states defined on the boundaries of this embedded patch.

To calculate the soft photon factor, we set $r^0=t$ and simplify the tail terms in the Green's function for $\vec{x} >> \vec{r}\left(\sigma\right)$
\begin{align}
\tilde{G}_M\left(\omega,\vec{x},\vec{r}\right)&=\lim_{\vec{x}\gg\vec{r}} G_M\left(\omega,\vec{x},\vec{r}\right) =\frac{i M}{16 \pi \omega} \left[\ln\left(\frac{\vert\vec{r}\vert+ \hat{n}.\vec{r}}{R}\right) + \int_{|\vec{r}|+\hat{n}.\vec{r}}^\infty\frac{du}{u} e^{i\omega u}\right] \frac{e^{i\omega (R - \hat{n}.\vec{r})}}{R}\,,\\\tilde{G}_l\left(\omega,\vec{x},\vec{r}\right)&=\lim_{\vec{x}\gg\vec{r}} G_l\left(\omega,\vec{x},\vec{r}\right)=-\frac{4i}{\omega} e^{i\omega (R - \hat{n}.\vec{r})} \hat{n}.\vec{r}\,.
\label{gfn.lim}
\end{align} 
We can then rewrite $\tilde{a}_i(\omega, \vec{x})$ in the large $\vert t \vert$ limit as 
\begin{align}
\tilde{a}_i(\omega, \vec{x})&=\tilde{a}_i^{(1)}(\omega, \vec{x})+\tilde{a}_i^{(2)}(\omega, \vec{x})+\tilde{a}_i^{(3)}(\omega, \vec{x})+\tilde{a}_i^{(4)}(\omega, \vec{x})+\tilde{a}_i^{(5)}(\omega, \vec{x}) + \tilde{a}_i^{(6)}(\omega, \vec{x})\notag\\& \qquad\qquad+ \tilde{a}_i^{(7)}(\omega, \vec{x})+  \tilde{a}_i^{(8)}(\omega, \vec{x}) + \tilde{a}_i^{(9)}(\omega, \vec{x}) + \tilde{a}_i^{(10)}(\omega, \vec{x})\,,
\end{align}
where the individual terms present in the above equation are defined as below:
\begin{align}
\tilde{a}_{i}^{(1)}\left(\omega,\vec{x}\right) = \frac{q}{1+\frac{x^2}{8l^2}}\frac{e^{i \omega R}}{4 \pi R} \int dt~ e^{i\omega(t-\hat{n}.\vec{r})} \frac{v_i }{1+ \phi(\vec{r})+\frac{r^2}{8l^2}} \,,\label{aai.1}
\end{align}
\begin{align}
\tilde{a}_i^{(2)}(\omega\,,\vec{x}) &= \frac{q}{2\pi}\int dt\, e^{i\omega t}\,v_k\nabla_k\nabla_i\tilde{G}_M\left(\omega,\vec{x},\vec{r}\right)\,, \label{aai.2}
\end{align}
\begin{align}
\tilde{a}_i^{(3)}(\omega\,,\vec{x}) &= \frac{i q}{\pi} \omega \int dt\, e^{i\omega t} \,\nabla_i\tilde{G}_M\left(\omega,\vec{x},\vec{r}\right)\,,\label{aai.3}
\end{align}
\begin{align}
\tilde{a}_i^{(4)}(\omega\,,\vec{x}) &=-\frac{i M q}{16\pi^2}\frac{e^{i\omega R}}{R} \omega \int dt v_i\bigg\{
\ln \frac{|\vec{r}\,'|+\hat{n}.\vec{r}\,'}{ R} \, e^{i\omega (t - \hat{n}.\vec{r}\,')} + \int_{|\vec{r}\,'|+\hat{n}.\vec{r}\,'}^\infty \frac{du}{u} e^{i\omega (t - \hat{n}.\vec{r}\,'+u)}\bigg\} \,,\label{aai.4}
\end{align}
\begin{align}
\tilde{a}_i^{(5)}(\omega\,,\vec{x}) &=-\frac{q}{4\pi}\int dt\, {e^{i\omega t}}\, v_i\nabla_k\nabla_k\tilde{G}_M\left(\omega,\vec{x},\vec{r}\right)\,,\label{aai.5}
\end{align}
\begin{align}
\tilde{a}_i^{(6)}(\omega\,,\vec{x}) &= \frac{4 Q m}{M} \int dt\, \frac{dt}{d\sigma}\, e^{i\omega t} \,v_k \nabla_k\nabla_i\tilde{G}_M\left(\omega,\vec{x},\vec{r}\right)\,, \label{aai.6}
\end{align}
\begin{align}
\tilde{a}_i^{(7)}(\omega\,,\vec{x}) &= -\frac{i 4 Q m}{M} \omega \int dt\, \frac{dt}{d\sigma}\, e^{i\omega t} \,v_i v_k\nabla_k\tilde{G}_M\left(\omega,\vec{x},\vec{r}\right)\,, \label{aai.7}
\end{align}
\begin{align}
\tilde{a}_i^{(8)}(\omega\,,\vec{x}) &= -\frac{i 2 Q m}{3 M} \omega \int dt\, \frac{dt}{d\sigma}\, e^{i\omega t} \,\nabla_i\tilde{G}_M\left(\omega,\vec{x},\vec{r}\right)\,, \label{aai.8}
\end{align}
\begin{align}
\tilde{a}_i^{(9)}(\omega\,,\vec{x}) &= \frac{i 2 Q m}{M} \omega \int dt\, \frac{dt}{d\sigma}\, e^{i\omega t} \, \vec{v}^2 \nabla_i\tilde{G}_M\left(\omega,\vec{x},\vec{r}\right)\,, \label{aai.9}
\end{align}
\begin{align}
\tilde{a}_i^{(10)}(\omega\,,\vec{x}) &=-\frac{q}{8\pi l^2}\omega^2\int dt\, {e^{i\omega t}} v_i \tilde{G}_l\left(\omega,\vec{x},\vec{r}\right)\,,\label{aai.10}
\end{align}
\begin{align}
\tilde{a}_i^{(11)}(\omega\,,\vec{x}) &=\frac{i q}{8\pi l^2} \omega\int dt\, {e^{i\omega t}}\nabla_i \tilde{G}_l\left(\omega,\vec{x},\vec{r}\right)\,,\label{aai.11}
\end{align}
\begin{align}
\tilde{a}_i^{(12)}(\omega\,,\vec{x}) &=-\frac{q}{32\pi l^2}\int dt\, {e^{i\omega t}}v_i \nabla_k\nabla_k\tilde{G}_l\left(\omega,\vec{x},\vec{r}\right)\,,\label{aai.12}
\end{align}
\begin{align}
\tilde{a}_i^{(13)}(\omega\,,\vec{x}) &=\frac{q}{16\pi l^2}\int dt\, {e^{i\omega t}}v_k \nabla_k\nabla_i\tilde{G}_l\left(\omega,\vec{x},\vec{r}\right)\,.\label{aai.13}
\end{align}

We can think of $\tilde{a}_i^{(2)}$ to $\tilde{a}_i^{(5)}$ as contributions due to the scatterer black hole's mass, $\tilde{a}_i^{(6)}$ to $\tilde{a}_i^{(9)}$ arising due to the black hole's charge and $\tilde{a}_i^{(10)}$ to $\tilde{a}_i^{(13)}$ due to the AdS potential. The soft limit evaluation of $\tilde{a}_i^{(2)}$ to $\tilde{a}_i^{(9)}$ has been previously derived in \cite{Fernandes:2020tsq}. Here we will evaluate $\tilde{a}_i^{(1)}$ in the soft limit. This term will give us the leading $1/l^2$ correction to the soft photon factor of asymptotically flat spacetime due to the AdS spacetime.

To evaluate the soft limit of $\tilde{a}_i^{(1)}$ we will be using the following relation
\begin{equation}
e^{i\omega(t-\hat{n}.\vec{r}(t))}=\frac{1}{i\omega}\frac{1}{\partial_t(t-\hat{n}.\vec{r}(t))}\frac{d}{dt}e^{i\omega(t-\hat{n}.\vec{r}(t))}=\frac{1}{i\omega}\frac{1}{(1-\hat{n}.\vec{v}(t))}\frac{d}{dt}e^{i\omega(t-\hat{n}.\vec{r}(t))} \,.\label{eas}
\end{equation} 
Using \ref{eas} and carrying out an integration by parts, we have
\begin{align}
\tilde{a}_{i}^{(1)}\left(\omega,\vec{x}\right) = -\frac{q}{1+\frac{x^2}{8l^2}}~\frac{e^{i \omega R}}{4 \pi R}~\frac{1}{i\omega} \int dt~ e^{i\omega(t-\hat{n}.\vec{r})} \frac{d}{dt}\left[\frac{v_i}{(1-\hat{n}.\vec{v}(t))}\frac{1}{1+ \phi(\vec{r})+\frac{r^2}{8l^2}}\right]\,.\label{kaan}
\end{align}
Since we assume $\frac{r^2}{l^2}\ll 1$ throughout our calculation, this implies that we have the following approximation
$$\frac{1}{1+ \phi(\vec{r})+\frac{r^2}{8l^2}}=\frac{1}{1+ \phi(\vec{r})}\left(1-\frac{r^2}{8l^2}\right)\,.$$
Hence \ref{kaan} simplifies to
\begin{align}
\tilde{a}_{i}^{(1)}\left(\omega,\vec{x}\right)& = -\frac{q}{1+\frac{x^2}{8l^2}}~\frac{e^{i \omega R}}{4 \pi R}~\frac{1}{i\omega} \int dt~ e^{i\omega(t-\hat{n}.\vec{r})} \frac{d}{dt}\left[\frac{v_i}{(1-\hat{n}.\vec{v}(t))}\frac{1}{1+ \phi(\vec{r})}\right]\,\notag\\&+ \frac{q}{8l^2}~\frac{e^{i \omega R}}{4 \pi R}~\frac{1}{i\omega} \int dt~ e^{i\omega(t-\hat{n}.\vec{r})} \frac{d}{dt}\left[\frac{v_i}{(1-\hat{n}.\vec{v}(t))}r^2\right]\notag\\&={\cal{A}}_1+{\cal{A}}_2+{\cal{A}}_3+\mathcal{O}(1/l^3,\phi^2)\,,
\end{align}
where
\begin{align}
{\cal{A}}_1&=-q~\frac{e^{i \omega R}}{4 \pi R}~\frac{1}{i\omega} \int dt~ e^{i\omega(t-\hat{n}.\vec{r})} \frac{d}{dt}\left[\frac{v_i}{(1-\hat{n}.\vec{v}(t))}\frac{1}{1+ \phi(\vec{r})}\right]\,,\\
{\cal{A}}_2&=q\frac{x^2}{8l^2}~\frac{e^{i \omega R}}{4 \pi R}~\frac{1}{i\omega} \int dt~ e^{i\omega(t-\hat{n}.\vec{r})} \frac{d}{dt}\left[\frac{v_i}{(1-\hat{n}.\vec{v}(t))}\right]\,,\label{kala2}\\
{\cal{A}}_3&=\frac{q}{8l^2}~\frac{e^{i \omega R}}{4 \pi R}~\frac{1}{i\omega} \int dt~ e^{i\omega(t-\hat{n}.\vec{r})} \frac{d}{dt}\left[\frac{v_i}{(1-\hat{n}.\vec{v}(t))}r^2\right]\,.\label{A}
\end{align}
Note that we are keeping terms up to leading order in $\phi$ and $1/l^2$. In the asymptotic expansion $\phi$ will take the form
\begin{equation}
\phi(\vec{r}(t)) = -\frac{M}{8 \pi \vert \vec{r}(t) \vert} = \mp \frac{M}{8 \pi \vert \vec{\beta}_{\pm} \vert t}   + \mathcal{O}\left(t^{-2}\right)\,, \label{as.pot}
\end{equation}
where we have used the parametrization of $r$ from \ref{asymrv}. ${\cal{A}}_1$ is the contribution on asymptotically flat spacetimes and has been evaluated in \cite{Fernandes:2020tsq}. Therefore we will concentrate on the other two integrals. To evaluate ${\cal{A}}_2$, we will substitute $v$ from \ref{asymrv} and perform the following expansion
\begin{equation}
v_i\left(1 - \hat{n}.\vec{v}\right)^{-1} = \left(1 - \hat{n}.\vec{\beta}_{\pm}\right)^{-1} \beta_{\pm i}\left[1 - \frac{C_{\pm}}{t} \frac{1}{1 - \hat{n}.\vec{\beta}_{\pm}}\right] + \mathcal{O}\left(t^{-2}\right) \,.\label{as.nv}
\end{equation}
Therefore \ref{kala2} reduces to
\begin{align}
{\cal{A}}_2&=q\frac{x^2}{8l^2}~\frac{e^{i \omega R}}{4 \pi R}~\frac{1}{i\omega} \int dt~ e^{-i\omega \tilde{g}(t)} \frac{d}{dt}\left[\frac{\beta_{\pm i}}{\left(1 - \hat{n}.\vec{\beta}_{\pm}\right)}\left(1 - \frac{C_{\pm}}{t} \frac{1}{1 - \hat{n}.\vec{\beta}_{\pm}}\right) \right] \,, \label{A2}\\
& \text{with} \quad \tilde{g}(t) = - (1-\hat{n}.\vec{\beta}_{\pm})t  - C_{\pm}\hat{n}.\vec{\beta}\ln t \,. \label{gtild.exp} 
\end{align}
In the $\omega\to 0$ limit, we have the following integral \cite{Laddha:2018myi}
\begin{align}
I_1 = \frac{1}{\omega} \int \limits_{-\infty}^{+\infty} dt e^{-i \omega g(t)} \frac{d}{dt}f(t) = \frac{1}{\omega} \left(f_{+} - f_{-}\right) + i \left(a_{+}k_{+} - a_{-}k_{-}\right) \text{ln} \omega^{-1} + \text{finite}\,,\label{int}
\end{align}
where
\begin{align}
f(t) &\to f_{\pm} + \frac{k_{\pm}}{t} \,, \qquad g(t) \to a_{\pm} t + b_{\pm} \ln \vert t \vert \,. \label{f}
\end{align}
Using \ref{int}, we find that the integral ${\cal{A}}_2$ in \ref{A2} evaluates to
\begin{align}
{\cal{A}}_2=& -\frac{i}{\omega}\frac{qx^2}{8l^2}\frac{e^{i \omega R}}{4 \pi R}\left[\frac{\beta_{+i}}{1 - \hat{n}.\vec{\beta}_{+}}-\frac{\beta_{-i}}{1 - \hat{n}.\vec{\beta}_{-}}\right]\notag\\&\qquad +\frac{qx^2}{8l^2}\frac{e^{i \omega R}}{4 \pi R}\ln \omega^{-1}\left[\frac{\beta_{+i}}{1 - \hat{n}.\vec{\beta}_{+}}C_+-\frac{\beta_{-i}}{1 - \hat{n}.\vec{\beta}_{-}}C_-\right] \,.\label{bit}
\end{align}
To consider the soft limit, we will need to take the simultaneous limits  $\omega\rightarrow 0$ and $l\rightarrow \infty$, while keeping $\omega l$ fixed. Therefore, by defining $\omega l = \gamma$, we find that \ref{bit} can be written as
\begin{align}
{\cal{A}}_2=& -i\frac{qx^2}{8\gamma^2}\frac{e^{i \omega R}}{4 \pi R}\omega\left[\frac{\beta_{+i}}{1 - \hat{n}.\vec{\beta}_{+}}-\frac{\beta_{-i}}{1 - \hat{n}.\vec{\beta}_{-}}\right]\notag\\&\qquad +\frac{qx^2}{8\gamma^2}\frac{e^{i \omega R}}{4 \pi R}\omega^2\ln \omega^{-1}\left[\frac{\beta_{+i}}{1 - \hat{n}.\vec{\beta}_{+}}C_+-\frac{\beta_{-i}}{1 - \hat{n}.\vec{\beta}_{-}}C_-\right]\,.
\end{align}
We thus note that in the $\omega\rightarrow 0$ limit, the ${\cal{A}}_2$ term is a finite contribution and does not provide the divergent terms in the soft factor. It follows that divergent contributions due to $\frac{1}{l^{2}}$ corrections in the soft factor will arise from those integrals which at least fall off like $\omega^{-3}$. 

The integrand in ${\cal{A}}_3$ involves the term $\frac{v_i}{(1-\hat{n}.\vec{v}(t))}r^2$, which has the following expansion in $t$ 
\begin{align} 
\frac{v_i}{(1-\hat{n}.\vec{v}(t))}r^2 &=\frac{\beta_{\pm i}}{1 - \hat{n}.\vec{\beta}_{\pm}} \vec{\beta}_{\pm}^2 \left[t^2-2 C_{\pm}~t\ln \vert t\vert-\frac{C_{\pm}}{1 - \hat{n}.\vec{\beta}_{\pm}}t+\frac{2C^2_{\pm}}{1 - \hat{n}.\vec{\beta}_{\pm}}\ln\vert t \vert\right] \notag\\ & \qquad \qquad \qquad + \frac{\beta_{\pm i}}{\left(1 - \hat{n}.\vec{\beta}_{\pm}\right)^3} C_{\pm}^2\vec{\beta}_{\pm}^2 \hat{n}.\vec{\beta}_{\pm} \left[1 - \frac{C_{\pm}\hat{n}.\vec{\beta}_{\pm}}{1 - \hat{n}.\vec{\beta}_{\pm}}\frac{1}{t}  \right] + \cdots \,,\label{exp}
\end{align}
where $\cdots$ in \ref{exp} refer to subleading terms that are $\mathcal{O}\left(\frac{1}{t^{2}}\right)$ and $\mathcal{O}\left(\frac{\ln \vert t \vert}{t}\right)$, which will always lead to finite terms in the soft factor and are hence ignored. The terms in the second line of \ref{exp} can be directly substituted into \ref{A} to provide the following contribution, which we denote as ${\cal{A}}^{(0)}_3$

\begin{align}
{\cal{A}}_3^{(0)}&= -\frac{i q}{8\gamma^2}~\frac{e^{i \omega R}}{4 \pi R}~\omega \int dt~ e^{- i\omega \tilde{g}(t)} \frac{d}{dt}\left[\frac{C_{\pm}^2\vec{\beta}_{\pm}^2 \hat{n}.\vec{\beta}_{\pm} \beta_{\pm i}}{\left(1 - \hat{n}.\vec{\beta}_{\pm}\right)^3} \left[1 - \frac{C_{\pm}\hat{n}.\vec{\beta}_{\pm}}{1 - \hat{n}.\vec{\beta}_{\pm}}\frac{1}{t}\right]\right]\,.
\label{a30}
\end{align}

In taking the simultaneous limits $\omega \to 0$ and $l \to \infty$ of the ${\cal{A}}_3^{(0)}$ integral in \ref{a30}, the overall coefficient ensures that we will have a non-divergent contribution in $\omega$ to the soft factor. 

To determine the other possible contributions of ${\cal{A}}_3$, we need to differentiate \ref{exp} with respect to $t$ and retain terms up to $1/t$. This gives us
\begin{align}
&\frac{d}{dt}\left[\frac{v_i}{(1-\hat{n}.\vec{v}(t))}r^2\right]= \frac{\beta_{\pm i}}{1 - \hat{n}.\vec{\beta}_{\pm}}\vec{\beta}_{\pm}^2\left[2 t - 2 C_{\pm}~\ln \vert t\vert -C_{\pm}\frac{3 - 2 \hat{n}.\vec{\beta}_{\pm}}{1 - \hat{n}.\vec{\beta}_{\pm}}+\frac{2 C^2_{\pm}}{1 - \hat{n}.\vec{\beta}_{\pm}}\frac{1}{t}\right]\,.
\label{diff.t}
\end{align}

To apply the integral \ref{int} we have to perform another integration by parts using \ref{eas} until we get the form of \ref{f}. By using the expression in \ref{diff.t} and subsequently performing the integration by parts, we find
\begin{align}
{\cal{A}}_3&=-\frac{q}{8l^2}~\frac{e^{i \omega R}}{4 \pi R}~\left(\frac{1}{i\omega}\right)^2 \int dt~ e^{-i\omega\tilde{g}(t)} \frac{d}{dt}\left[\frac{1}{(1-\hat{n}.\vec{v}(t))}\frac{\beta_{\pm i}}{1 - \hat{n}.\vec{\beta}_{\pm}} \vec{\beta}_{\pm}^2\right.\notag\\&\qquad \qquad \qquad \qquad \qquad\left.\left(2 t - 2 C_{\pm}~\ln \vert t\vert - C_{\pm}\frac{3 - 2 \hat{n}.\vec{\beta}_{\pm}}{1 - \hat{n}.\vec{\beta}_{\pm}}+\frac{2 C^2_{\pm}}{1 - \hat{n}.\vec{\beta}_{\pm}}\frac{1}{t}\right)\right]\notag\\&=-\frac{q}{8l^2}~\frac{e^{i \omega R}}{4 \pi R}~\left(\frac{1}{i\omega}\right)^2 \int dt~ e^{-i\omega\tilde{g}(t)} \frac{d}{dt}\left[\frac{\beta_{\pm i}}{(1 - \hat{n}.\vec{\beta}_{\pm})^2}\vec{\beta}_{\pm}^2 \left(2 t - 2 C_{\pm}~\ln \vert t\vert\right.\right.\notag\\&\left.\left. \quad \qquad \qquad \qquad \qquad \qquad \qquad +\frac{1}{t}\left(\frac{C_{\pm}}{1 - \hat{n}.\vec{\beta}_{\pm}}\right)^2\left(2 + \hat{n}.\vec{\beta}_{\pm}\right)- 3 \frac{C_{\pm}}{1 - \hat{n}.\vec{\beta}_{\pm}}\right)\right] \,. \label{a3bp1}
\end{align}

We will now separate the ${\cal{A}}_3$ integral above into terms which can be evaluated using \ref{int} and those which require an additional integration by parts. The terms in the parenthesis of \ref{a3bp1} which are constant and proportional to $\frac{1}{t}$ provide the following contribution which we denote as ${\cal{A}}_3^{(1)}$

\begin{align}
{\cal{A}}_3^{(1)}&=-\frac{q}{8\gamma^2}~\frac{e^{i \omega R}}{4 \pi R} \int dt~ e^{-i\omega\tilde{g}(t)}\frac{\beta_{\pm i}}{(1 - \hat{n}.\vec{\beta}_{\pm})^2} \frac{d}{dt}\left[3 \frac{C_{\pm}}{1 - \hat{n}.\vec{\beta}_{\pm}}-\frac{1}{t}\left(\frac{C_{\pm}}{1 - \hat{n}.\vec{\beta}_{\pm}}\right)^2\left(2 + \hat{n}.\vec{\beta}_{\pm}\right)\right]\,.
\end{align}

Evaluating this integral in the soft limit using \ref{int}, we find a finite contribution that is irrelevant to the soft photon factor. The remaining terms in the integrand of \ref{a3bp1} require another differentiation followed by an integration by parts. This contribution, which we denote by ${\cal{A}}_3^{(2)}$, takes the form

\begin{align}
{\cal{A}}_3^{(2)}&=-\frac{q}{8l^2}~\frac{e^{i \omega R}}{4 \pi R}~\left(\frac{1}{i\omega}\right)^2 \int dt~ e^{-i\omega\tilde{g}(t)} \frac{\beta_{\pm i}}{(1 - \hat{n}.\vec{\beta}_{\pm})^2}\left(2\vec{\beta}_{\pm}^2-2\vec{\beta}_{\pm}^2 \frac{C_{\pm}}{t}\right)\notag\\&=\frac{iq}{8\gamma^2}~\frac{e^{i \omega R}}{4 \pi R}~\frac{1}{\omega} \int dt~ e^{-i\omega\tilde{g}(t)} \frac{\beta_{\pm i}}{(1 - \hat{n}.\vec{\beta}_{\pm})^3}\frac{d}{dt}\left(2\vec{\beta}_{\pm}^2-2\vec{\beta}_{\pm}^2\frac{C_{\pm}}{t}\frac{1}{1 - \hat{n}.\vec{\beta}_{\pm}}\right)\,.
\end{align}
This integral in the soft limit evaluates to
\begin{align}
{\cal{A}}_3^{(2)}&=\frac{1}{\gamma^2}\left[\frac{iq}{4\omega}~\frac{e^{i \omega R}}{4 \pi R}~\Bigg\{\frac{\beta_{+ i}}{(1 - \hat{n}.\vec{\beta}_{+})^3}\vec{\beta}_{+}^2-\frac{\beta_{- i}}{(1 - \hat{n}.\vec{\beta}_{-})^3}\vec{\beta}_{-}^2\Bigg\}\right.\notag\\&\left.-\frac{q}{4}\ln\omega^{-1}~\frac{e^{i \omega R}}{4 \pi R}~\Bigg\{\frac{\beta_{+ i}}{(1 - \hat{n}.\vec{\beta}_{+})^3}C_{+}\vec{\beta}_{+}^2-\frac{\beta_{- i}}{(1 - \hat{n}.\vec{\beta}_{-})^3}C_{-}\vec{\beta}_{-}^2\Bigg\}\right]\,.\label{main}
\end{align}
The divergent terms in this result contribute to the soft photon factor. From \ref{main} we can write the soft factor following \ref{sfa}. 

The other radiative field terms, those from $\tilde{a}_i^{(10)}$ to $\tilde{a}_i^{(13)}$, are either finite in the soft limit or are proportional to $k_i$. Gauge invariance can be used to show that terms which go like $k_i$ do not contribute to the soft factor. The invariance of $S_{\text{em}}$ in \ref{sfa} under $\epsilon^{\mu} \to \epsilon^{\mu} + k^{\mu}$ imposes the constraint $k^{\mu}\tilde{a}_{\mu} = 0$.  This implies that $\tilde{a}_{0}$ can be determined from $\tilde{a}_i$ and allows us to set $\epsilon^0 = 0$.  In addition, the expressions for $\tilde{a}_{\mu}$ are determined only up to a choice in gauge. Denoting the arbitrary gauge parameters by $\lambda$, we have the following gauge transformations
\begin{equation}
 \delta \tilde{a}_{\mu} = \lambda k_{\mu} \,. 
\label{emgr.gtf}
\end{equation}
Using \ref{emgr.gtf} in \ref{sfa}, we thus have $ k_{\mu}\epsilon^{\mu} = 0 $. This provides the condition on the polarization vector as $k_i \epsilon^i =0$. 
 
Therefore the $1/l^2$ contribution to the soft photon factor arises only from \ref{main}. On substituting in \ref{sfa} we find
\begin{align}
S_{\rm em}^l=&\frac{q}{4\gamma^2 }\epsilon^i\left[-\frac{1}{\omega}~~\Bigg\{\frac{\beta_{+ i}}{(1 - \hat{n}.\vec{\beta}_{+})^3}\vec{\beta}^2_{+}-\frac{\beta_{- i}}{(1 - \hat{n}.\vec{\beta}_{-})^3}\vec{\beta}^2_{-}\Bigg\}\right.\notag\\&\left.- i \ln\omega^{-1}~\Bigg\{\frac{C_+\beta_{+ i}}{(1 - \hat{n}.\vec{\beta}_{+})^3}\vec{\beta}^2_{+} - \frac{C_-\beta_{- i}}{(1 - \hat{n}.\vec{\beta}_{-})^3}\vec{\beta}^2_{-}\Bigg\}\right]\,.
\label{softads.em}
\end{align}
We can describe the above result in terms of the momentum and angular momentum of the probe particle, and the momentum of the soft photon. From \ref{asymrv} we find the asymptotic momenta (as $t \to \infty$) to have the expressions
\begin{equation}
    p_{(1)} = \frac{m}{\sqrt{1 - \vec{\beta}_-^2}}\left(1,\vec{\beta}_-\right) \,, \qquad p_{(2)} = -\frac{m}{\sqrt{1 - \vec{\beta}_+^2}}\left(1,\vec{\beta}_+\right)\,.
    \label{asym.mom}
\end{equation}
The overall negative sign in $p_{(2)}$ reflects the convention for outgoing momenta. Using \ref{asymrv} we also have for the position four vector
\begin{equation}
     r_{(1)} = \left(t,\vec{\beta}_-t - C_-\vec{\beta}_- \ln\vert t\vert\right) \,, \qquad r_{(2)} = \left(t,\vec{\beta}_+t - C_+\vec{\beta}_+ \ln\vert t\vert\right) \,,
\end{equation}
with the coefficients of the logarithmic terms in the position vector, $c^{\mu}_{(a)}$, taking the form
\begin{equation}
    c_{(1)} = \left(1, - C_- \vec{\beta}_-\right) \,, \qquad c_{(2)} = \left(1, - C_+ \vec{\beta}_+\right)\,.
\end{equation}
Here for simplicity we are ignoring an overall constant term in the position vector, which will not contribute to the soft factor result (as it is a finite contribution in the $t \to \infty$ limit). 

The angular momentum of the probe particle can be written as
\begin{equation}
    j^{\mu \nu}_{(a)} = r^{\mu}_{(a)}p^{\nu}_{(a)} - r^{\nu}_{(a)}p^{\mu}_{(a)}+\text{spin}=(c_a^{\mu}p_{a}^{\nu}-c_a^{\nu}p_{a}^{\mu})\ln|t|+\cdots\,,
\end{equation}
and thus has the following non-vanishing components
\begin{equation}
    j_{(1)}^{0i} = \frac{m}{\sqrt{1 - \vec{\beta}_-^2}} C_-\vec{\beta}_- \ln\vert t\vert \,, \quad j_{(2)}^{0i} = -\frac{m}{\sqrt{1 - \vec{\beta}_+^2}} C_+\vec{\beta}_+ \ln\vert t\vert\,.
\label{asym.amom}
\end{equation}

Lastly, the outgoing soft photon has the momentum $k = -\omega(1,\hat{n})$. Using this along with the expressions in \ref{asym.mom} and \ref{asym.amom}, and replacing $\ln|t|$ by $\ln\omega^{-1}$, we determine that \ref{softads.em} can be written as the following sum
\begin{align}
S_{\rm em}^l &= S_{\rm em}^{l\,(0)} + S_{\rm em}^{l\,(1)}  \,, 
\end{align}
where $S_{\rm em}^{l\,(0)}$ and $S_{\rm em}^{l\,(1)}$ are the universal leading and subleading contributions respectively on AdS spacetimes, with the expressions
\begin{align}
S_{\rm em}^{l\, (0)} &= \frac{q}{4l^2} \sum_{a=1}^2 \left(-1\right)^{a-1} \frac{\epsilon_{\mu} p^{\mu}_{(a)}}{p_{(a)}.k} \frac{\vec{p}^2_{(a)}}{\left(p_{(a)}.k\right)^2} \,, \label{sphf}\\
S_{\rm em}^{l\,(1)} &= i \frac{q}{4l^2} \sum_{a=1}^2 \left(-1\right)^{a-1} \frac{\epsilon_{\nu} k_\rho j^{\rho \nu}_{(a)}}{p_{(a)}.k} \frac{\vec{p}^2_{(a)}}{\left(p_{(a)}.k\right)^2}\notag\\ &=i\frac{q}{4l^2}\ln \omega^{-1}\sum_{a=1}^2 \left(-1\right)^{a-1} \frac{\epsilon_{\nu} k_\rho (c_a^{\rho}p_a^{\nu}-c_a^{\nu}p_a^{\rho})}{p_{(a)}.k}\frac{\vec{p}^2_{(a)}}{\left(p_{(a)}.k\right)^2}\,.
\label{sphf.sub}
\end{align}
Apart from the factor $\frac{\vec{p}^2_{(a)}}{\left(p_{(a)}.k\right)^2}$ appearing in each of the above sums, the expressions are those for the universal leading and subleading soft factors on asymptotically flat spacetimes, in the  case of the scattering of a single probe particle. The summation index values $a = 1$ and $a=2$ correspond to the outgoing and incoming configuration of the probe, respectively. 

It can be noted that the momentum soft factor expression is not covariant, due to the involvement of $\vec{p}_{(a)}^2$. This is a consequence of working in the `static gauge' choice for the polarization vector of the electromagnetic field. We would always be required to adopt some gauge in the probe scattering calculations on a fixed curved spacetime, which is needed to derive the AdS result. Furthermore, we see from the dependence of additional momenta factors that the above result on AdS spacetimes cannot be represented as an overall phase of the asymptotically flat spacetime result. 

We can also note that the gauge transformation $\epsilon^{\mu} \to \epsilon^{\mu} + k^{\mu}$ in \ref{sphf} does not lead to the manifest vanishing of these factors unlike on asymptotically flat spacetimes. Rather, the requirement of the resulting sums over the momenta becomes a statement of momentum conservation for scattering processes up to $1/l^2$ corrections on asymptotically AdS spacetimes. Hence the $1/l^2$ corrected soft factor results are gauge invariant for all viable scattering processes on the spacetime.

For completeness, we note the soft photon factor that can be derived on asymptotically flat backgrounds arising from contributions in ${\cal{A}}_1$, $\tilde{a}_i^{(3)}$, $\tilde{a}_i^{(4)}$ and $\tilde{a}_i^{(5)}$ \cite{Fernandes:2020tsq}
\begin{align}
&S_{\text{em}}^{\text{flat}} =
 - \frac{q}{\omega} \left(\frac{\vec{\epsilon}. \vec{\beta}_+}{1 - \hat{n}.\vec{\beta}_+} - \frac{\vec{\epsilon}. \vec{\beta}_-}{1 - \hat{n}.\vec{\beta}_-}\right) - i q \ln \omega^{-1} \left(\frac{C_+}{1 - \hat{n}.\vec{\beta}_+}\vec{\epsilon}. \vec{\beta}_+ - \frac{C_-}{1 - \hat{n}.\vec{\beta}_-}\vec{\epsilon}. \vec{\beta}_-\right) \notag\\
&\qquad  \qquad - i q \frac{M}{4 \pi} \ln\left(\omega R\right) \left(\frac{\vec{\epsilon}. \vec{\beta}_+}{1-\hat{n}.\vec{\beta}_+} - \frac{\vec{\epsilon}. \vec{\beta}_-}{1-\hat{n}.\vec{\beta}_-}\right)\,.
\label{ai.soft1}
\end{align} 

In this case, the soft factor can be expressed in covariant form as given in \cite{Laddha:2018myi}.

\section{Classical soft graviton factor}\label{sec5}
In this section, we will simply present the results on the contribution of the charge and mass of the scatterer black hole to the soft graviton factor. For a detailed calculation we refer the reader to Appendix \ref{SGT}. The soft graviton factor can be split in two parts as,
\begin{equation}
S_{\text{gr}}(\epsilon, k)=S_{\rm gr}^{\rm flat}(\epsilon, k) +S_{\rm gr}^{l}(\epsilon, k)\,, 
\end{equation}
where $S_{\rm gr}^{\text{flat}}(\epsilon, k)$ is the contribution due to the black hole mass and charge, which was derived in \cite{Fernandes:2020tsq}. The result in its covariant form can be expressed as \cite{Laddha:2018myi}
\begin{align}
S_{\rm gr}^{\rm flat}(\epsilon, k)=& \sum_{a=1}^2 {\epsilon_{\mu\nu} p_{(a)}^\mu p_{(a)}^\nu\over p_{(a)}.k} +
 i\, \sum_{a=1}^2 
{\epsilon_{\mu\nu} p_{(a)}^\mu k_\rho \over p_{(a)}. k} j^{\rho\nu}_{(a)}
\,.
\label{gr_p}
\end{align}
In  \ref{gr_p}, $p_{(1)}$ and $p_{(2)}$ are the  momenta of the probe particle before and after the scattering, while $j_{(1)}$ and $j_{(2)}$ are the angular momenta of the probe before and after the scattering. As in the previous section, we choose the convention that all momenta and angular momenta are positive for ingoing and negative for outgoing particles. The indices $\mu,\nu,\cdots$ run over all spacetime coordinates in \ref{gr_p}. The covariant expression has an implicit dependence on the masses and charges in the scattering process. The explicit form of the above equation on replacing the momenta and angular momenta has been provided in \ref{SGT}.

Further in \cite{Banerjee:2020dww}, we derived the leading order contribution of the cosmological constant to the soft graviton factor on the AdS Schwarzschild black hole spacetime. Since to our order of approximation, the metric in the isotropic coordinates does not receive any correction from the charge of the black hole, the results for the soft factor remains the same. Here we present this result in covariant form, in terms of the incoming and outgoing momenta of the probe particle
\begin{align}
S_{\rm gr}^l &= S_{\rm gr}^{l\,(0)} + S_{\rm gr}^{l\,(1)} \,, \qquad \text{with} \notag\\
S_{\rm gr}^{l\,(0)} &= \frac{1}{2l^2} \sum_{a=1}^2 \frac{\epsilon_{\mu \nu} p^{\mu}_{(a)}p^{\nu}_{(a)}}{p_{(a)}.k} \frac{\vec{p}^2_{(a)}}{\left(p_{(a)}.k\right)^2} \left( 3 + \frac{\vec{p}^2_{(a)}}{p^2_{(a)}}\right)\,, \notag\\
S_{\rm gr}^{l\,(1)} &= i \frac{1}{2l^2} \sum_{a=1}^2 \frac{\epsilon_{\mu \nu} p^{\mu}_{(a)}k_\rho j^{\rho \nu}_{(a)}}{p_{(a)}.k} \frac{\vec{p}^2_{(a)}}{\left(p_{(a)}.k\right)^2} \left( 3 + \frac{\vec{p}^2_{(a)}}{p^2_{(a)}}\right)\,,
\label{sgrf}
\end{align}
where $S_{\rm gr}^{l\,(0)}$ and $S_{\rm gr}^{l\,(1)}$ are respectively the universal leading and subleading contributions to the soft graviton factor on AdS spacetimes.
Note that for the four momentum we have $p^2_{(a)} = -m^2$. The leading and subleading soft factor sums are, as in the soft photon results in \ref{sphf} and \ref{sphf.sub}, expressed in terms of a product of the flat spacetime result and other momentum dependent terms. The gauge invariance followes for all physical scattering process, as discussed in the last section.

\section{Ward identities from soft photon factors} \label{sec6}
In the previous sections we derived classical soft photon and graviton factors in an asymptotically AdS theory, to leading order $1/l^2$ in the large AdS radius limit. Considering the large AdS radius limit is important for computational simplifications as well as for defining a proper soft limit in AdS space. Our analysis can be interpreted as $1/l^2$ corrections of classical soft theorems in asymptotically flat spacetime in large AdS radius limit. Thus we expect these AdS corrected soft factors derived in \ref{sphf} and \ref{sgrf} to satisfy similar relations as the usual flat spacetime classical soft factors. By now, it is a known fact that classical soft factors partially reproduce the quantum soft factors. These results hold exactly at the level of tree and all loop scattering amplitudes for the leading soft factor \footnote{for subleading and sub-subleading parts, the quantum soft factor reduces to the classical one in the classical limit \cite{Sahoo:2020ryf}}. On the other hand, it is also known that soft theorems with the leading quantum soft factor contribution manifest as certain Ward identities arising from large gauge transformations on asymptotically flat spacetimes. We can thus anticipate that soft theorems involving leading soft factor terms, with AdS corrections, should have a corresponding realization in a Ward identity. 

In this section, we concern ourselves with the soft photon case. We will more specifically be interested in the equivalence of large gauge Ward identity with the soft photon theorem in the case of a massless scattering process. While the soft factor expressions derived in previous sections resulted from a massive probe particle scattering process, the universality of the tree level soft photon factor enables us to consider its expression in any scattering process involving soft particles. The analysis in particular involves the derivation of an expression for the soft photon number mode from a saddle point approximation. To maintain a flow of the paper and to establish notations and definitions, we first review the correspondence between the large gauge Ward identity and leading soft factor for a theory on asymptotically flat spacetimes \cite{He:2014cra, He:2014laa, Strominger:2017zoo} to the leading order and then derive the same for our case \footnote{Experts may skip this section and directly go to section \ref{WIA}}.

For a residual gauge parameter $\varepsilon(z\,,\bar{z})$ at null infinity which satisfies the antipodal boundary condition 
\begin{equation}
\varepsilon(z\,,\bar{z}) \vert_{\mathscr{I}^+_-} = \varepsilon(z\,,\bar{z}) \vert_{\mathscr{I}^-_+} \,,
\end{equation}
we can define the total charge for the Maxwell field on $\mathscr{I}^+$ and $\mathscr{I}^-$ in terms of boundary contributions on asymptotically flat spacetimes

\begin{equation}
Q_{\varepsilon}^+ =  \int_{\mathscr{I}^+_-} \varepsilon * F \,, \qquad Q_{\varepsilon}^- =  \int_{\mathscr{I}^-_+} \varepsilon * F  \,.
\label{scharge}
\end{equation}

here $(z, \bar z)$ are coordinates on the Celestial 2-sphere. The above charges satisfy the following conservation equation

\begin{equation}
Q_{\varepsilon}^+ -  Q_{\varepsilon}^- = 0\,.
\label{ccon}
\end{equation}

We can now consider Maxwell's equations in the presence of a source to re-express the charges in \ref{scharge} as the following integrals over $\mathscr{I}^+$ and $\mathscr{I}^-$

\begin{align}
Q_{\varepsilon}^+ = - \int_{\mathscr{I}^+} du d^2z \left(\partial_z \varepsilon F^{(0)}_{u \bar{z}} + \partial_{\bar{z}} \varepsilon F^{(0)}_{u z} \right) + \int_{\mathscr{I}^+} du d^2z \varepsilon \gamma_{z\bar{z}} j_{u}^{(2)} \,,\label{maxu.eq}\\
Q_{\varepsilon}^- = -  \int_{\mathscr{I}^-} dv d^2z \left(\partial_z \varepsilon F^{(0)}_{v \bar{z}} + \partial_{\bar{z}} \varepsilon F^{(0)}_{v z} \right) + \int_{\mathscr{I}^-} dv d^2z \varepsilon \gamma_{z\bar{z}} j_{v}^{(2)} \,.\label{maxv.eq}
\end{align}

In the above expressions $\gamma_{z\bar{z}}$ is the metric on the $2$-sphere and the superscripts on the field strength tensors and currents represent the order of the coefficient in their corresponding $r^{-1}$ expansions in asymptotically flat spacetimes. The first integrals in $Q_{\varepsilon}^+$ and $Q_{\varepsilon}^-$ are known as the soft charges. The field strength tensors involved in the soft charges can be expressed in terms of their soft modes. Further in the absence of asymptotic magnetic fields and magnetic monopoles, we have the following condition

\begin{equation}
F_{z \bar{z}}\vert_{\mathscr{I}^{\pm}_{\pm}} = 0\,.
\end{equation}

In this case, it can be shown that we can define the soft modes of the field strength tensors in terms of the following real scalars $\mathcal{N}^{\pm}$ that satisfy

\begin{align}
\partial_z \mathcal{N}^{+}(z\,,\bar{z}) = \int_{\mathscr{I}^+} du F^{(0)}_{u z} \,, \qquad & \partial_{\bar{z}} \mathcal{N}^{+}(z\,,\bar{z}) = \int_{\mathscr{I}^+} du F^{(0)}_{u \bar{z}}\,, \notag\\
\partial_z \mathcal{N}^{-}(z\,,\bar{z}) = \int_{\mathscr{I}^-} dv F^{(0)}_{v z} \,, \qquad & \partial_{\bar{z}} \mathcal{N}^{-}(z\,,\bar{z}) = \int_{\mathscr{I}^-} dv F^{(0)}_{v \bar{z}}.
\label{smode.fl}
\end{align}

Using \ref{smode.fl} in \ref{maxu.eq} and \ref{maxv.eq} then gives us the total charge expressions

\begin{align}
Q_{\varepsilon}^+ = 2 \int d^2z \mathcal{N}^{+} \partial_{z}\partial_{\bar{z}}\varepsilon + \int_{\mathscr{I}^+} du d^2z \varepsilon \gamma_{z\bar{z}} j_{u}^{(2)} \,,\label{maxu2.eq}\\
Q_{\varepsilon}^- = 2 \int d^2z \mathcal{N}^{-} \partial_{z}\partial_{\bar{z}}\varepsilon + \int_{\mathscr{I}^-} dv d^2z \varepsilon \gamma_{z\bar{z}} j_{v}^{(2)} \,.\label{maxv2.eq}
\end{align}

With these classical results, we can now describe the Ward identity. Given a scattering process going from an incoming state $\vert \text{in}\rangle$ on $\mathscr{I}^-$ to a state $\vert \text{out}\rangle$ on $\mathscr{I}^+$ governed by an $S$-matrix $\mathcal{S}$, the charge conservation in \ref{ccon} now takes the form

\begin{equation}
\langle \text{out} \vert \hat{Q}^{+}_{\varepsilon} \mathcal{S} - \mathcal{S} \hat{Q}^-_{\varepsilon} \vert \text{in}\rangle = 0 \,,
\label{scon.q}
\end{equation}

where $\hat{Q}^{\pm}_{\varepsilon}$ are operator versions of the expressions in \ref{maxu2.eq} and \ref{maxv2.eq}. The integrated hard charges can be expressed as the following sum over hard charges in the incoming and outgoing states

\begin{align}
\int_{\mathscr{I}^+} du d^2z \varepsilon \gamma_{z\bar{z}} j_{u}^{(2)} = \sum_{k= \text{out}}Q_k  \varepsilon \left(z_k\,,\bar{z}_k\right) \,, \label{hard.out} \\
\int_{\mathscr{I}^-} dv d^2z \varepsilon \gamma_{z\bar{z}} j_{v}^{(2)} = \sum_{k= \text{in}}Q_k  \varepsilon \left(z_k\,,\bar{z}_k\right) \,. \label{hard.in}
\end{align}

The index `$k$' run over incoming and outgoing hard particles in the respective sums, while the coordinates $\{z_k\,,\bar{z}_k\}$ denote their asymptotic positions on the Celestial $2$-sphere. Using the above expressions in \ref{maxu2.eq} and \ref{maxv2.eq}, we find that the operators $\hat{Q}^{\pm}_{\varepsilon}$ take the form
\begin{align}
\langle \text{out} \vert \hat{Q}_{\varepsilon}^+ = 2 \int d^2z \partial_{z}\partial_{\bar{z}}\varepsilon \langle \text{out} \vert\mathcal{N}^{+}  + \sum_{k= \text{out}}Q_k  \varepsilon\left(z_k\,,\bar{z}_k\right) \langle \text{out} \vert \,,\label{maxuop.eq}\\
\hat{Q}_{\varepsilon}^- \vert \text{in}\rangle = 2 \int d^2z \partial_{z}\partial_{\bar{z}}\varepsilon \mathcal{N}^{-} \vert \text{in}\rangle + \sum_{k= \text{in}}Q_k \varepsilon\left(z_k\,,\bar{z}_k\right) \vert \text{in}\rangle \,.\label{maxvop.eq}
\end{align}

Substituting \ref{maxuop.eq} and \ref{maxvop.eq} in \ref{scon.q} then gives us the following Ward identity~\cite{He:2014cra}
\begin{align}
&2 \int d^2z \partial_z \partial_{\bar{z}} \varepsilon\left(z\,,\bar{z}\right) \langle \text{out} \vert \mathcal{N}^{+} \left(z\,,\bar{z}\right) \mathcal{S} - \mathcal{S} \mathcal{N}^{-} \left(z\,,\bar{z}\right) \vert \text{in}\rangle \notag\\
&\quad = \left[\sum_{k= \text{in}}Q_k \varepsilon\left(z_k\,,\bar{z}_k\right) - \sum_{k= \text{out}}Q_k  \varepsilon\left(z_k\,,\bar{z}_k\right)\right]  \langle \text{out} \vert \mathcal{S}\vert \text{in}\rangle \,.
\label{ph.wardt}
\end{align}

%
We can express the left hand side of \ref{ph.wardt} entirely in terms of either the soft photon mode $\cal{N}^+$ or $\cal{N}^-$. This follows from the CPT invariance of matrix elements involving the in and out soft photons. This implies in particular that 
$$\langle \text{out} \vert \mathcal{N}^{+} \left(z\,,\bar{z}\right) \mathcal{S}\vert \text{in}\rangle = - \langle \text{out} \vert \mathcal{S} \mathcal{N}^{-} \left(z\,,\bar{z}\right)\vert \text{in}\rangle \,.$$
Hence \ref{ph.wardt} can be expressed as
\begin{align}
&4 \int d^2z \partial_z \partial_{\bar{z}} \varepsilon\left(z\,,\bar{z}\right) \langle \text{out} \vert \mathcal{N}^{+} \left(z\,,\bar{z}\right) \mathcal{S}  \vert \text{in}\rangle \notag\\
&\quad = \left[\sum_{k= \text{in}}Q_k \varepsilon\left(z_k\,,\bar{z}_k\right) - \sum_{k= \text{out}}Q_k  \varepsilon\left(z_k\,,\bar{z}_k\right)\right]  \langle \text{out} \vert \mathcal{S}\vert \text{in}\rangle \,.
\label{ph.ward}
\end{align}

To simplify the notation going forward, we will refer to the soft photon mode $\mathcal{N}^+$ as $\mathcal{N}$ and its associated outgoing operators $a_{+}^{\text{out}}(\omega \hat{x})$ and $a_{-}^{\text{out}}(\omega \hat{x})^{\dagger}$ as $a_{+}^{\text{flat}}(\omega \hat{x})$ and $a_{-}^{\text{flat}}(\omega \hat{x})^{\dagger}$ respectively. An expression for $\partial_{z}\mathcal{N}$ can be formally derived from a mode expansion of the gauge fields and a saddle point approximation~\cite{Strominger:2017zoo}

\begin{equation}
\partial_{z}\mathcal{N} =  -\frac{1}{8\pi}\frac{\sqrt{2}}{1+z\bar{z}} \lim_{\omega \to 0}\left[\omega a_{+}^{\text{flat}}(\omega \hat{x}) + \omega a_{-}^{\text{flat}}(\omega \hat{x})^{\dagger}\right] \,.
\label{dn.fl}
\end{equation}

We next consider the gauge parameter. A particularly convenient choice as considered in \cite{Strominger:2017zoo} for $\varepsilon\left(z\,,\bar{z}\right)$ is the following

\begin{equation}
\varepsilon\left(z\,,\bar{z}\right) = \frac{1}{z- z_k} \,.
\label{g.fl}
\end{equation}

In particular, the derivative of \ref{g.fl} satisfies

\begin{equation}
\partial_{\bar{w}}\varepsilon(z\,,\bar{z}) = 2\pi \delta^{(2)}\left(z - w\right) \,.
\label{g.delt}
\end{equation}

By substituting \ref{g.delt} and \ref{dn.fl} in \ref{ph.ward}, we find

\begin{equation}
\lim_{\omega\to 0}\langle \text{out}\vert \omega a^{\text{flat}}_{+}(\omega \hat{x}) \mathcal{S} \vert \text{in} \rangle = \frac{(1 + z \bar{z})}{\sqrt{2}} \left[\sum_{k=\text{out}} \frac{Q_k}{z - z_{k}} - \sum_{k=\text{in}} \frac{Q_k}{z - z_{k}}\right] \langle \text{out}\vert \mathcal{S} \vert \text{in} \rangle  \,.
\label{ward.fin}
\end{equation}

The large gauge Ward identity is formally derived as a property satisfied by quantized fields of a given theory. However, we may also derive the Ward identity using the soft photon theorem. We recall that the soft photon theorem relates a scattering process with a soft photon insertion to the scattering process without the soft photon by a soft factor. Assuming for simplicity the insertion of the soft photon in the `out' state, the theorem provides the relation
\begin{align}
\langle \text{out}\vert a^{\text{flat}}_{+}(\omega \hat{x})\mathcal{S} \vert \text{in} \rangle  = S^{\text{flat}}_{\text{em}} \langle \text{out}\vert \mathcal{S} \vert \text{in} \rangle \,,
\label{spt}
\end{align}

where $a^{\text{flat}}_{+}$ denotes the creation operator in the outgoing state with helicity `$+$' and $S^{\text{flat}}_{\text{em}}$ is taken to be the leading contribution to the soft photon factor

\begin{equation}
    S^{\text{flat}\;(0)}_{\text{em}} =  \left[\sum_{k= \text{out}} \frac{p_{(k)}.\epsilon_+}{p_{(k)}.q} Q_k - \sum_{k=\text{in}} \frac{p_{(k)}.\epsilon_+}{p_{(k)}.q} Q_k\right] \,,
\label{spf.lead}
\end{equation}

and $\mathcal{S}$ is a generic scattering process in flat spacetime. Our analysis in previous sections considered a single probe particle with charge $q$. Thus in this case, we set $Q_k=q$ in \ref{spf.lead}.

We can now readily derive the Ward identity in \ref{ph.ward} from \ref{spt}. For this derivation, we only need to use the expressions for the momenta of all the particles (hard and soft) and the polarization vector for the soft particle. For the soft particle and its polarization, we have the following conditions 

\begin{equation}
q_{\mu}q^{\mu} = 0 \,, \quad q_{\mu} \epsilon^{\pm \mu}(\vec{q}) = 0 \,, \quad \epsilon^{\mu}_{\alpha}\epsilon^*_{\beta \mu} = \delta_{\alpha \beta}\,,
\label{polq.cond}
\end{equation}

where the indices $\alpha \,,\beta$ represent the polarization directions $\pm$. A natural choice for the null vector $q^{\mu}$ is the following
\begin{equation}
q^{\mu} = \omega \left(1 \,, {\vec{x} \over r}\right) = \frac{\omega}{1+ z \bar{z}} \left( 1+ z \bar{z}\,, z+\bar{z}\,,-i(z - \bar{z})\,, 1- z \bar{z}\right) \,,
\label{q.ch}
\end{equation}

with $\omega$ the frequency of the soft particle. From this choice of $q^{\mu}$ one can determine the polarization vectors

\begin{equation}
\epsilon^{+\mu} = \frac{1}{\sqrt{2}}\left(\bar{z} \,, 1\,, -i\,,-\bar{z}\right) \,, \quad \epsilon^{-\mu} = \frac{1}{\sqrt{2}}\left(z \,, 1\,, i\,,-z\right)\,.
\label{pol.ch}
\end{equation}

Since the hard particles are also massless, but of finite energy, using \ref{q.ch} we can choose
\begin{equation}
p_{k}^{\mu} = \frac{E_{\vec{p}}}{1+ z \bar{z}} \left( 1+ z \bar{z}\,, z+\bar{z}\,,-i(z - \bar{z})\,, 1- z \bar{z}\right)\,.
\label{p.ch}
\end{equation}

If we now substitute \ref{q.ch}, \ref{pol.ch} and \ref{p.ch} in \ref{spf.lead}, we find
\begin{align}
 S^{\text{flat}\,(0)}_{\text{em}} &=  \left[\sum_{k= \text{out}} \frac{p_{(k)}.\epsilon_+}{p_{(k)}.q} Q_k - \sum_{k=\text{in}} \frac{p_{(k)}.\epsilon_+}{p_{(k)}.q} Q_k\right] \notag\\
& = \frac{1 + z \bar{z}}{\sqrt{2} \omega}  \left[\sum_{k=\text{out}} \frac{1}{z - z_{k}}Q_k - \sum_{k=\text{in}} \frac{1}{z - z_{k}}Q_k\right]\,.
\label{spf}
\end{align} 

Thus on substituting \ref{spf} in \ref{spt}, we get
\begin{equation}
\lim_{\omega\to 0}\frac{\sqrt{2}}{1 + z \bar{z}}\langle \text{out}\vert \omega a^{\text{flat}}_{+}(\omega \hat{x}) \mathcal{S} \vert \text{in} \rangle =  \left[\sum_{k=\text{out}} \frac{1}{z - z_{k}}Q_k - \sum_{k=\text{in}} \frac{1}{z - z_{k}}Q_k\right] \langle \text{out}\vert \mathcal{S} \vert \text{in} \rangle\,.
\label{spt2}
\end{equation}

This result agrees with the Ward identity in \ref{ward.fin}.


%

In the presence of AdS corrections of asymptotically flat spacetimes, we can determine the tree level soft theorem which involves the soft factor derived from our classical scattering analysis. However, a first principles derivation of the Ward identity using quantized fields is not particularly clear. As previously described, the reason for this is that the $\gamma^{-2}$ correction to the asymptotically flat spacetime soft factor arises from a double scaling limit. This implies that unlike asymptotically flat spacetimes, the $\gamma^{-2}$ correction cannot result from a unique saddle point approximation of the gauge fields. Rather, there can exist several contributions to $\gamma = l \omega$ as we simultaneously take $l\to \infty$ and $\omega \to 0$.  Thus, while the saddle point approximation determines the soft mode $\mathcal{N}$ in terms of creation and annihilation operators on asymptotically flat spacetimes, this does not extend to possible $1/l^2$ corrections arising from the nearly flat limit of asymptotically AdS spacetimes. These corrections could be determined by assuming that the equivalence of tree level soft theorems and large gauge Ward identities on asymptotically flat spacetimes continues to hold under small corrections. As we will show in the following, this is highly constraining on the expressions for the $1/l^2$ corrections of the soft photon mode and residual gauge parameter.

\subsection{$1/l^2$ correction of the large gauge Ward identity} \label{WIA}

We derived the leading soft photon factor correction $S^{l \, (0)}_{\text{em}}$ from a classical scattering analysis on an AdS spacetime in \ref{sphf}. To interpret the corresponding soft theorem for a massless scattering process in terms of an associated Ward identity, we must define the asymptotic surface on which it is defined. Our scattering process with the assumptions on a large impact parameter is well approximated by the large $l$ limit of a global AdS spacetime, with the black hole as a point particle source located at the origin. The global AdS$_4$ metric is described by the metric
\begin{equation}
    ds^2 = \frac{l^2}{\cos^2 \rho}\left[-d\tau^2 + d\rho^2 + \sin^2 \rho d\Omega_2^2\right] \,.
    \label{ads.met}
\end{equation}
In the limit of $l \to \infty$, we can identify an asymptotically flat patch close to the origin of the AdS spacetime \cite{Hijano:2019qmi} that has recently been used to demonstrate the equivalence of the asymptotically flat spacetime large gauge Ward identity with a CFT Ward identity~\cite{Hijano:2020szl}. In our case, the effect of $1/l^2$ corrections will be shown to affect the soft photon mode. Hence the effect of $1/l^2$ corrections in the soft photon case has a natural interpretation as a perturbation of the asymptotically flat spacetime large gauge Ward identity, evaluated across null infinity of the patch embedded in the larger AdS spacetime. This is represented in Fig.1,

\begin{figure}[h]
\begin{center}
\includegraphics[scale=0.4]{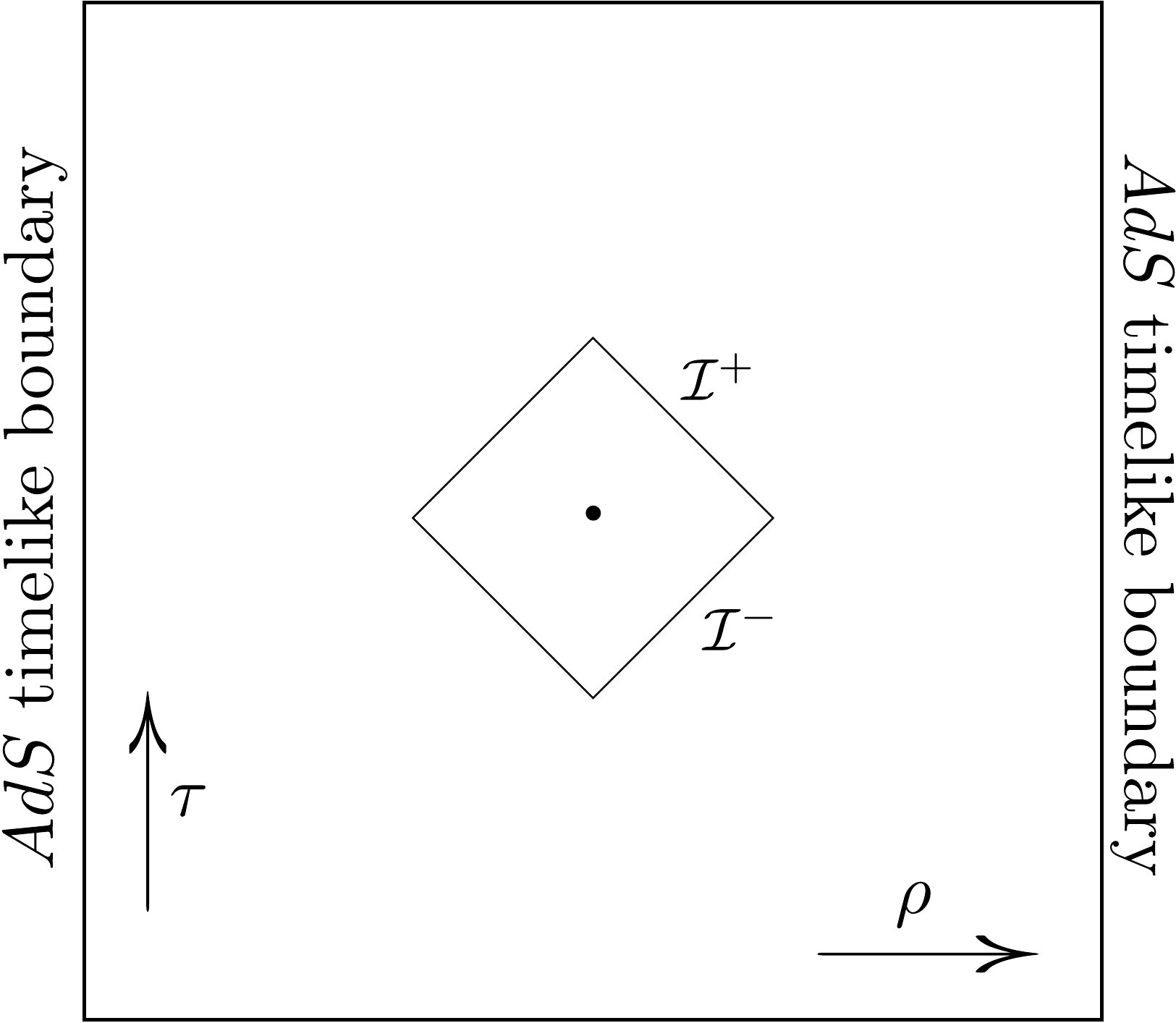}
\end{center}
\caption{The Ward identity is defined across the null surfaces of the asymptotically flat spacetime patch embedded within a larger AdS spacetime. The black dot at the center represents the black hole treated as a point particle in our scattering approximation. The effect of $1/l^2$ corrections is to perturb the asymptotically flat spacetime Ward identity defined on the surface of the patch.}
\end{figure}

As noted through the expression in \ref{tlr}, the $1/l^2$ corrections do not affect the flat spacetime asymptotic trajectories of particles in any scattering process that provide the soft factor expressions up to subleading order. Hence the parametrizations for the momenta of the hard (and soft) massless particles and the polarization can be taken to be the same as those on asymptotically flat spacetime given in \ref{q.ch}, \ref{pol.ch} and \ref{p.ch}.

Using these expressions in \ref{sphf}, we have the the following result for the soft factor correction in terms of the asymptotic coordinates of the particles $(z\,,\bar{z})$, $(z_k\,,\bar{z}_k)$ and the soft photon frequency

\begin{align}
S^{l \, (0)}_{\text{em}} &= \frac{1}{4 l^2} \left[\sum_{k= \text{out}} \frac{p_{(k)}.\epsilon_+}{p_{(k)}.q}\frac{\vec{p}^2_{(k)}}{\left(p_{(k)}.q\right)^2} Q_k - \sum_{k=\text{in}} \frac{p_{(k)}.\epsilon_+}{p_{(k)}.q}\frac{\vec{p}^2_{(k)}}{\left(p_{(k)}.q\right)^2}Q_k\right] \notag\\
& =  \frac{\left(1 + z \bar{z}\right)^3}{16 \sqrt{2} \gamma ^2 \omega}\left[\sum_{k=\text{out}} \frac{\left(1 + z_k \bar{z}_k\right)^2}{(\bar{z}-\bar{z}_k)^2}\frac{1}{(z-z_k)^3}Q_k - \sum_{k=\text{in}} \frac{\left(1 + z_k \bar{z}_k\right)^2}{(\bar{z}-\bar{z}_k)^2}\frac{1}{(z-z_k)^3}Q_k\right]\,,
\label{spf.ads}
\end{align} 

where we have denoted $l^2 \omega^2$ as $\gamma^2$. 

A consistent deformation of the soft photon theorem given in \ref{spt} on asymptotically flat spacetimes requires $1/l^{2}$ corrections on either side of the equation. This leads to the following identity for scattering processes 

\begin{align}
\langle \text{out}\vert a^{\text{flat}}_{+}(\omega \hat{x})\mathcal{S} \vert \text{in} \rangle + \langle \text{out}\vert \tilde{a}_{+}(\omega \hat{x})\mathcal{S} \vert \text{in} \rangle  = \left(S^{\text{flat}\,(0)}_{\text{em}} + S^{l \, (0)}_{\text{em}}\right) \langle \text{out}\vert \mathcal{S} \vert \text{in} \rangle\,,
\label{spt.ads}
\end{align}

with a separate theorem satisfied by the corrected components
\begin{align}
 \langle \text{out}\vert \tilde{a}_{+}(\omega \hat{x})\mathcal{S} \vert \text{in} \rangle  =  S^{l \, (0)}_{\text{em}} \langle \text{out}\vert \mathcal{S} \vert \text{in} \rangle\,.
 \label{ads.th}
\end{align}
In \ref{spt.ads}, $S^{\text{flat}\,(0)}_{\text{em}}$ and $S^{l\,(0)}_{\text{em}}$ are the universal leading contributions without and with $1/l^2$ corrections respectively, and $\tilde{a}_+$ denotes a creation operator in the out state corresponding to the $1/l^2$ correction. Note that the $\tilde{a}_+$ operator must be distinct from the flat space soft photon operator and in particular has a different dimension. We will make this point clear by defining $\tilde{a}_+ = \frac{1}{l^{2}}a_+^{\text{AdS}}$ later on.

The way to interpret \ref{spt.ads} is that there exists a scattering process in flat spacetime, and we insert a soft photon mode which involves a $1/l^2$ contribution. This leads to a corresponding $1/l^2$ correction of the soft factor. This result informs us that there must be $1/l^2$ corrections of both $\partial_w \mathcal{N}$ and the gauge parameter $\varepsilon$ appearing in the usual flat spacetime Ward identity given in \ref{ph.ward}. The reason for $1/l^2$ corrections in $\partial_w \mathcal{N}$ also follows from $1/l^2$ corrections of the soft photon operator.

On the other hand, the soft factor informs us on the $1/l^2$ correction of the gauge parameter $\varepsilon(z_k\,,\bar{z}_k)$. This must involve the terms appearing in the parenthesis of  \ref{spf.ads}. However, we desire that the gauge parameter have some of the properties of the asymptotically flat spacetime gauge parameter given in \ref{g.fl}. We accordingly define the total gauge parameter to be

\begin{align}
\varepsilon(w\,,\bar{w}) &= \varepsilon^{\text{flat}}(w\,,\bar{w}) + \frac{1}{l^2}\varepsilon^{\text{AdS}}(w\,,\bar{w})\,; \notag\\
\varepsilon^{\text{flat}}(w\,,\bar{w}) & = \frac{1}{\omega - z}\,,\notag\\
\varepsilon^{\text{AdS}}(w\,,\bar{w}) & = \frac{(1 + z \bar{z})^2}{\left(\bar{w} - \bar{z}\right)^2} \frac{(1 + w \bar{w})^2}{\left( w - z\right)^3} \notag\\
&= \varepsilon^{\text{flat}}(w\,,\bar{w}) \frac{(1 + z \bar{z})^2}{\left(\bar{w} - \bar{z}\right)^2} \frac{(1 + w \bar{w})^2}{\left( w - z\right)^2}\,.
\label{ep.ads}
\end{align}

We thus have defined $\varepsilon^{\text{AdS}} = \Omega~ \varepsilon^{\text{flat}}$, where $\Omega = \frac{(1 + z \bar{z})^2}{\left(\bar{w} - \bar{z}\right)^2} \frac{(1 + w \bar{w})^2}{\left( w - z\right)^2}$ is a factor invariant under the interchange of $w \leftrightarrow z$ and $\bar{w} \leftrightarrow \bar{z}$. We note that this property ensures that $\varepsilon^{\text{AdS}}$ does change sign under the interchange of $w$ with $z$, as in the case of $\varepsilon^{\text{flat}}$. This requirement led us to the include the factor $(1 + w \bar{w})^2$ in the definition of $\varepsilon^{\text{AdS}}$ apart from the term appearing in the parenthesis of \ref{spf.ads}. 

As noted above, we also have $1/l^2$ corrections of the soft photon creation and annihilation operators. We accordingly have a correction of $\partial_z \mathcal{N}$
\begin{align}
\partial_w \mathcal{N} & = \partial_w \mathcal{N}^{\text{flat}} + \frac{1}{l^2}\partial_w \mathcal{N}^{\text{AdS}}\,; \label{dzn.ads}\\
\partial_w \mathcal{N}^{\text{flat}}& = - \frac{\sqrt{2}}{8 \pi} \frac{1}{1 + w \bar{w}} \lim_{\omega \to 0} \left[\omega a_+^{\text{flat}}(\omega \hat{x}) + \omega a_-^{\text{flat}}(\omega \hat{x})^{\dagger}\right]\,. 
\label{dzn.flat}
\end{align}

The goal is to determine the form of $\partial_z \mathcal{N}^{\text{AdS}}$ such that $1/l^2$ corrections of the Ward identity for large gauge transformations in \ref{ph.ward} gives the soft photon theorem in \ref{spt.ads}, with the factor given in \ref{spf.ads}. By using the expressions from \ref{ep.ads} and \ref{dzn.ads} in \ref{ph.ward}, and collecting the $1/l^2$ coefficient, we find

\begin{align}
4 &\int d^2 w \left[\partial_{\bar{w}}\varepsilon^{\text{flat}}(w\,,\bar{w})  \langle \text{out}\vert \partial_{w}\mathcal{N}^{\text{AdS}}(w\,,\bar{w}) \mathcal{S} \vert \text{in} \rangle + \partial_{\bar{w}}\varepsilon^{\text{AdS}}(w\,,\bar{w})  \langle \text{out}\vert \partial_{w}\mathcal{N}^{\text{flat}}(w\,,\bar{w}) \mathcal{S} \vert \text{in} \rangle\right]   \notag\\
&= \left[\sum_{k=\text{in}} \frac{\left(1 + z_k \bar{z}_k\right)^2}{(\bar{z}-\bar{z}_k)^2}\frac{\left(1 + z \bar{z}\right)^2}{(z-z_k)^3} Q_k - \sum_{k=\text{out}} \frac{\left(1 + z_k \bar{z}_k\right)^2}{(\bar{z}-\bar{z}_k)^2}\frac{\left(1 + z \bar{z}\right)^2}{(z-z_k)^3} Q_k\right] \langle \text{out}\vert \mathcal{S} \vert \text{in} \rangle\,.
\label{exp.ward}
\end{align}

\ref{exp.ward} will agree with the correction to the soft photon theorem given in \ref{ads.th} if 

\begin{align}
\frac{4}{l^2} &\int d^2 w \left[\partial_{\bar{w}}\varepsilon^{\text{flat}}(w\,,\bar{w})  \langle \text{out}\vert \partial_{w}\mathcal{N}^{\text{AdS}}(w\,,\bar{w}) \mathcal{S} \vert \text{in} \rangle + \partial_{\bar{w}}\varepsilon^{\text{AdS}}(w\,,\bar{w})  \langle \text{out}\vert \partial_{w}\mathcal{N}^{\text{flat}}(w\,,\bar{w}) \mathcal{S} \vert \text{in} \rangle\right]   \notag\\
&= -\frac{16 \sqrt{2}}{ (1+ z\bar{z})} \lim_{\omega \to 0} \langle \text{out}\vert \omega^3 \tilde{a}_{+}(\omega \hat{x})\mathcal{S} \vert \text{in} \rangle \,.
\label{lhsexp.ward}
\end{align}

This can only be true if $\partial_w \mathcal{N}^{\text{AdS}}$ involves a sum of two parts -- a term $\partial_w \mathcal{N}^{\text{AdS}}_1$ which provides the $\tilde{a}_+$ contribution and another term $\partial_w \mathcal{N}^{\text{AdS}}_2$ which only involves $a^{\text{flat}}_+$. We hence assume

\begin{equation}
\partial_w \mathcal{N}^{\text{AdS}} = \partial_w \mathcal{N}^{\text{AdS}}_1 + \partial_w \mathcal{N}^{\text{AdS}}_2\,,
\label{dn.ads}
\end{equation}

with
\begin{align}
\partial_w \mathcal{N}^{\text{AdS}}_1 &= h(w\,, \bar{w}) \lim_{\omega \to 0} \left[\omega^3 a^{\text{AdS}}_+(\omega \hat{x}) + \omega^3 a^{\text{AdS}}_-(\omega \hat{x})^{\dagger}\right] \,,\label{dna1.p}\\
\partial_w \mathcal{N}^{\text{AdS}}_2 &= g(w\,, \bar{w}) \lim_{\omega \to 0} \left[\omega a_+^{\text{flat}}(\omega \hat{x}) + \omega a_-^{\text{flat}}(\omega \hat{x})^{\dagger}\right]\,,
\label{dna2.p}
\end{align}

where 
\begin{equation}
    \tilde{a}_{\pm}(\omega \hat{x}) = \frac{1}{l^2}a_{\pm}^{\text{AdS}}(\omega \hat{x}) \,, \quad \tilde{a}_{\pm}(\omega \hat{x})^{\dagger} = \frac{1}{l^2} a_{\pm}^{\text{AdS}}(\omega \hat{x})^{\dagger}\,.
\end{equation}

We can now determine the expressions for $h(w\,, \bar{w})$ and $g(w\,, \bar{w})$ from the LHS of \ref{exp.ward}, which can be expressed as
\begin{align} 
4 &\int d^2 w \left[\partial_{\bar{w}}\varepsilon^{\text{flat}}\langle \text{out}\vert \partial_{w}\mathcal{N}^{\text{AdS}}_1\mathcal{S} \vert \text{in} \rangle +  \partial_{\bar{w}}\varepsilon^{\text{flat}}\langle \text{out}\vert \partial_{w}\mathcal{N}^{\text{AdS}}_2\mathcal{S} \vert \text{in} \rangle + \partial_{\bar{w}}\varepsilon^{\text{AdS}} \langle \text{out}\vert \partial_{w}\mathcal{N}^{\text{flat}} \mathcal{S} \vert \text{in} \rangle\right] \notag\\
& = 4 \int d^2 w \partial_{\bar{w}}\varepsilon^{\text{flat}}\langle \text{out}\vert \partial_{w}\mathcal{N}^{\text{AdS}}_1\mathcal{S} \vert \text{in} \rangle  \notag\\
& \qquad \qquad \qquad - 4 \int d^2 w \left[\varepsilon^{\text{flat}}\langle \text{out}\vert \partial_{\bar{w}}\partial_{w}\mathcal{N}^{\text{AdS}}_2\mathcal{S} \vert \text{in} \rangle + \varepsilon^{\text{AdS}} \langle \text{out}\vert \partial_{\bar{w}}\partial_{w}\mathcal{N}^{\text{flat}} \mathcal{S} \vert \text{in} \rangle \right]\,.
\label{dzn.exp}
\end{align}

Hence the first term in the RHS of \ref{dzn.exp} gives us

\begin{equation}
4  \int d^2 w \partial_{\bar{w}}\varepsilon^{\text{flat}}\langle \text{out}\vert \partial_{w}\mathcal{N}^{\text{AdS}}_1\mathcal{S} \vert \text{in} \rangle  = 8 \pi \langle \text{out}\vert \partial_{z}\mathcal{N}^{\text{AdS}}_1 (z\,,\bar{z})\mathcal{S} \vert \text{in} \rangle\,.
\end{equation}

Using the expression from \ref{dna1.p}, we then find
\begin{align}
&h(z\,, \bar{z}) = -\frac{16 \sqrt{2}}{8 \pi (1 + z \bar{z})} \,, \notag\\
& \Rightarrow \partial_{z}\mathcal{N}^{\text{AdS}}_1 (z\,,\bar{z}) = -\frac{16 \sqrt{2}}{8 \pi (1 + z \bar{z})} \lim_{\omega \to 0} \left[\omega^3 a^{\text{AdS}}_+(\omega \hat{x}) + \omega^3 a^{\text{AdS}}_- (\omega \hat{x})^{\dagger}\right]\,.
\label{dna1}
\end{align}


We now also require that the last two terms in the RHS of \ref{dzn.exp} cancel, which determines the $\partial_{z}\mathcal{N}^{\text{AdS}}_2$ contribution. From taking the $\bar{w}$ derivative of $\partial_w \mathcal{N}^{\text{flat}}$ in \ref{dzn.flat} we have
 
\begin{equation}
\partial_{\bar{w}} \partial_w \mathcal{N}^{\text{flat}} =  \frac{\sqrt{2}}{8 \pi} \frac{w}{(1 + w \bar{w})^2} \lim_{\omega \to 0} \left[\omega a_+^{\text{flat}}(\omega \hat{x}) + \omega a_-^{\text{flat}}(\omega \hat{x})^{\dagger}\right]\,.
\label{ddnfl}
\end{equation}



Using \ref{ddnfl} and the expressions for $\varepsilon^{\text{AdS}}(w\,,\bar{w})$ and $\partial_w \mathcal{N}^{\text{AdS}}_2$ from \ref{ep.ads} and \ref{dna2.p} respectively, we then find that the terms in the last line of \ref{dzn.exp} to be

\begin{align}
\int d^2 w & \left[\varepsilon^{\text{flat}}\langle \text{out}\vert \partial_{\bar{w}}\partial_{w}\mathcal{N}^{\text{AdS}}_2\mathcal{S} \vert \text{in} \rangle + \varepsilon^{\text{AdS}} \langle \text{out}\vert \partial_{\bar{w}}\partial_{w}\mathcal{N}^{\text{flat}} \mathcal{S} \vert \text{in} \rangle \right]\notag\\
&= \int d^2 w \varepsilon^{\text{flat}} \left[\partial_{\bar{w}}g(w\,, \bar{w}) + \frac{\sqrt{2} }{8 \pi} \frac{(1 + z \bar{z})^2}{\left(\bar{w} - \bar{z}\right)^2} \frac{w}{\left( w - z\right)^2} \right]\langle \text{out}\vert \omega a_+^{\text{flat}}(\omega \hat{x}) \mathcal{S} \vert \text{in} \rangle\,.
\label{dzn2.exp}
\end{align}

Thus \ref{dzn2.exp} vanishes if

\begin{equation}
g(w\,, \bar{w}) = - \int d\bar{w} \frac{\sqrt{2}}{8 \pi} \frac{(1 + z \bar{z})^2}{\left(\bar{w} - \bar{z}\right)^2} \frac{w}{\left( w - z\right)^2} = \frac{\sqrt{2}}{8 \pi} \frac{(1 + z \bar{z})^2}{\left(\bar{w} - \bar{z}\right)} \frac{w}{\left( w - z\right)^2}\,.
\end{equation}



Hence from \ref{dna2.p} the result for $\partial_w \mathcal{N}^{\text{AdS}}_2$ takes the form

\begin{equation}
\partial_w \mathcal{N}^{\text{AdS}}_2 = \frac{\sqrt{2}}{8 \pi} \frac{(1 + z \bar{z})^2}{\left(\bar{w} - \bar{z}\right)} \frac{w}{\left( w - z\right)^2} \lim_{\omega \to 0} \left[\omega a_+^{\text{flat}}(\omega \hat{x}) + \omega a_-^{\text{flat}}(\omega \hat{x})^{\dagger}\right]\,.
\label{dna2}
\end{equation}

Thus the equation for the $1/l^2$ corrected Ward identity on asymptotically flat spacetimes is

\begin{align}
4 \int d^2w &\left[\partial_{\bar{w}}\varepsilon^{\text{flat}}(w\,,\bar{w})  \langle \text{out}\vert \partial_{w}\mathcal{N}^{\text{AdS}}(w\,,\bar{w}) \mathcal{S} \vert \text{in} \rangle + \partial_{\bar{w}}\varepsilon^{\text{AdS}}(w\,,\bar{w})  \langle \text{out}\vert \partial_{w}\mathcal{N}^{\text{flat}}(w\,,\bar{w}) \mathcal{S} \vert \text{in} \rangle\right]\notag\\
&\qquad \qquad = \left[\sum_{k=\text{in}} Q_k \varepsilon^{\text{AdS}}(z_{k}\,,\bar{z}_k) - \sum_{k=\text{out}} Q_k \varepsilon^{\text{AdS}}(z_{k}\,,\bar{z}_k)\right]  \langle \text{out}\vert \mathcal{S} \vert \text{in} \rangle \,, 
\label{ph.ward2}
\end{align}

which agrees with the corrected soft photon theorem in \ref{ads.th} by choosing $\varepsilon(w\,,\bar{w})$ as in \ref{ep.ads}, and $\partial_w \mathcal{N} (w\,, \bar{w})$ as in \ref{dn.ads}.

To summarize our main result in this section, we argued that $1/l^2$ corrections in the soft photon factor arise due to a perturbation of the soft photon theorem on asymptotically flat spacetimes. The perturbed soft theorem was given in \ref{ads.th}. Using the known equivalence between soft theorems and large gauge Ward identities on asymptotically flat spacetimes, we could then derive a perturbed Ward identity which is given in \ref{ph.ward2}. In the process of demonstrating this equivalence up to $1/l^2$, we determined the corrections of the gauge parameter in \ref{ep.ads} and the soft photon mode in \ref{dna1} and \ref{dna2}. Our results are defined across null infinity on the asymptotically flat patch that arises in the large $l$ limit of asymptotically AdS spacetimes, as indicated in Fig. 1.

\section{Conclusion and open questions}\label{sec7}

Defining a quantum soft theorem in asymptotically AdS spaces is not only a technically involved problem but also is an unclear issue as the notion of asymptotic in and out states are not well defined in AdS spacetime. Thus an alternative way to look for a possible soft factorization is required. The  analysis of \cite{Laddha:2018rle,Laddha:2018myi} for asymptotically flat theories showed that the soft factorization is also evident in classical radiation profiles. For asymptotically flat spacetimes, the radiative parts of the electromagnetic and gravitational fields produced in a classical scattering process provide the same leading quantum soft factor obtained from $S$-matrix, up to the usual gauge ambiguity. Therefore, we looked for a similar behaviour in asymptotically AdS systems with a small cosmological constant and found a similar factorization by assuming the cosmological constant as a perturbation parameter over asymptotically flat gravity. By considering the large impact parameter scattering of a probe particle with a black hole in AdS spacetime, we derived the $1/l^2$ corrections due to the AdS potential to the known leading $\omega^{-1}$ and subleading $\ln \omega^{-1}$ soft photon and soft graviton factors of four dimensional asymptotically flat spacetimes.

In \cite{Hijano:2020szl} and \cite{Hijano:2019qmi}, asymptotically flat spacetime scattering amplitudes were shown to result from the $l \to \infty$ limit of AdS boundary correlation functions.  The scattering amplitudes resulting from this limit are defined on the boundary of an asymptotically flat spacetime patch around the center of AdS. A description of bulk fields in terms of boundary operators, determined by using the HKLL formalism \cite{Hamilton:2006az}, can identify photon operators in the flat spacetime patch in terms of a $U(1)$ current. Additionally, the large gauge Ward identity for the soft photon theorem on asymptotically flat spacetimes was shown to be equivalent to a conformal Ward identity in taking the limit of $l \to \infty$ and $\omega \to 0$ simultaneously while keeping $\omega l$ fixed~\cite{Hijano:2020szl}. This precisely corresponds to the double scaling limit considered in this paper, and in a previous derivation of the soft graviton factor from classical scattering processes on asymptotically AdS spacetimes~\cite{Banerjee:2020dww}. 

Our analysis reveals that up to $1/l^2$ corrections, the double scaling limit does not affect hard particle trajectories from their flat spacetime expressions, while soft factors do involve corrections due to the consideration of an asymptotically AdS spacetime. This suggests that asymptotically flat spacetime large gauge Ward identities should be perturbed from $1/l^2$ corrections of the soft particles, while being defined on the same asymptotic boundaries for hard particles in the $l \to \infty$ of asymptotically AdS spacetimes, i.e the flat spacetime patch around the center. In Section \ref{sec6}, we made use of the universality of the leading soft photon factor to investigate $1/l^2$ corrections to the flat spacetime large gauge Ward identity in a massless scattering process. We determined that the perturbed identity results from $1/l^2$ corrections of the asymptotically flat spacetime soft photon mode and gauge parameter, which we derived up to an overall factor. This derivation only assumed that large gauge Ward identities and soft theorems on asymptotically flat spacetimes continue to remain equivalent under small perturbations of the soft particles.

It will be interesting to relate the $1/l^2$ corrected soft photon Ward identity with a CFT Ward identity on asymptotically AdS spacetimes. These results can be derived using the formalism in \cite{Hijano:2020szl} and \cite{Hijano:2019qmi}, while now retaining $1/l^{2}$ terms in the large AdS radius limit. The derivation lies outside the scope of this paper and will be presented in future work.  \vspace{2em}\\
{\bf Acknowledgements}\vspace{1em}\\
We would like to thank Sayali Bhatkar and Arindam Bhattacharjee for useful discussions. Our work is partially supported by a SERB ECR grant, GOVT of India. Finally, we thank the people of India for their generous support to the basic sciences.
\appendix
\section{Classical soft graviton theorem}\label{SGT}
In this section we will consider the contribution of the scatterer black hole charge to the gravitational soft factor. For a detailed calculation of the gravitation soft factor as a contribution of potential due to the black hole mass and AdS radius we refer the reader to \cite{Fernandes:2020tsq} and \cite{Banerjee:2020dww} respectively. 

$S_{\text{gr}}$ in $D=4$ dimensions can be computed using
\begin{equation}
S_{\text{gr}} = i \frac{4 \pi R}{e^{i \omega R}} \epsilon^{ij}\tilde{e}_{ij} (\omega, \vec{x}) \,.\label{sft4.gr}
\end{equation}
Similar to the the electromagnetic case, we let $r^0 = t$, assume $x >> r(t)$ and we will write \ref{hija.fin} as
\begin{align}
\tilde{e}_{ij}(\omega, \vec{x})& = \tilde{e}^{(1)}_{ij}(\omega, \vec{x}) + \tilde{e}^{(2)}_{ij}(\omega, \vec{x}) + \tilde{e}^{(3)}_{ij}(\omega, \vec{x})+ \tilde{e}^{(4)}_{ij}(\omega, \vec{x}) + \tilde{e}^{(5)}_{ij}(\omega, \vec{x}) + \tilde e^{(6)}_{ij}(\omega, \vec{x})+ \tilde e^{(7)}_{ij}(\omega, \vec{x})\notag\\&\qquad\qquad+\tilde e^{(8)}_{ij}(\omega, \vec{x})+\tilde e^{(9)}_{ij}(\omega, \vec{x})+\tilde e^{(10)}_{ij}(\omega, \vec{x})\,,
\end{align}
where
\begin{align} 
\tilde{e}^{(1)}_{ij}(\omega, \vec{x}) 
&=\frac{m~e^{i\omega R}}{ 4\, \pi\, R} \frac{1}{1-\frac{x^2}{2l^2}}\int \frac{dt}{\left(1+2\Phi(\vec{r}(t))+\frac{r^2}{4l^2}\right)}\frac{dt}{d\sigma}\, v_i v_j \, 
e^{i\omega (t-\hat{n}.\vec{r}(t))} +\text{boundary terms}
 \, ,\label{eij.1}\\
\tilde e^{(2)}_{ij}(\omega, \vec{x}) &= \frac{m}{2\pi} \int dt \frac{dt}{d\sigma}\, e^{i\omega t}(1+\vec{v}^2) \, \left(\nabla_i\nabla_j -\frac{1}{2} \delta_{ij} \, \nabla_k\nabla_k\right)\tilde{G}_M\left(\omega,\vec{x},\vec{r}\right)\,, \label{eij.2}\\
\tilde e^{(3)}_{ij}(\omega, \vec{x})&= -i\, \frac{M m}{16\, \pi^2} \, \omega \, \frac{e^{i\omega R}}{R}\int dt \frac{dt}{d\sigma}\, v_i\, v_j \,\notag\\&\hspace{5em}\bigg\{ \ln \frac{|\vec{r}\,'|+\hat{n}.\vec{r}\,'}{ R} \, e^{i\omega (t - \hat{n}.\vec{r}\,')}+ \int_{|\vec{r}\,'|+\hat{n}.\vec{r}\,'}^\infty \frac{du}{u} e^{i\omega (t - \hat{n}.\vec{r}\,'+u)}\bigg\}\,, \label{eij.3}\\
\tilde e^{(4)}_{ij}(\omega, \vec{x}) &= \frac{i\omega m}{\pi} \int dt \frac{dt}{d\sigma}\, e^{i\omega t}\left(v_i\nabla_j + v_j \nabla_i \right)\,\tilde{G}_M\left(\omega,\vec{x},\vec{r}\right)\,, \label{eij.4}\\ 
\tilde e^{(5)}_{ij}(\omega, \vec{x}) &= -i\omega\frac{qQ}{2\pi M}\int dt e^{i\omega t}\delta_{ij}v^k\nabla_k\tilde{G}_M\left(\omega,\vec{x},\vec{r}\right)\,, \label{eij.5}\\
\tilde e^{(6)}_{ij}(\omega, \vec{x}) &=i\frac{qQ}{\pi M} \int dte^{i\omega t}\left(\nabla_i\nabla_j -\frac{1}{2} \delta_{ij} \, \nabla_k\nabla_k\right)\tilde{G}_M\left(\omega,\vec{x},\vec{r}\right)\,,
\label{eij.6}\\
\tilde e^{(7)}_{ij}(\omega, \vec{x}) &= i\omega\frac{qQ}{2\pi M} \int dt e^{i\omega t}\left(v_i\nabla_j + v_j \nabla_i \right)\tilde{G}_M\left(\omega,\vec{x},\vec{r}\right)\,,\label{eij.7}
\end{align}
\begin{align}
\tilde e^{(8)}_{ij}(\omega, \vec{x}) =& -\frac{m}{8\pi l^2}\int dt \frac{dt}{d\sigma}\left[-~ \left(2-\vec{v}^2\right)\n_i \n_j - \frac{3}{2}\delta_{ij}v_k v_m\n_k\n_m- \frac{1}{2}\delta_{ij}v^2\n_k\n_k\right.\notag\\&\left.- ~\frac{3}{2}\left(v_k v_j\n_k\n_i+v_k v_i\n_k\n_j\right)+v_iv_j\n_k\n_k\right]\tilde{G}_l\left(\omega,\vec{x},\vec{r}\right)\,,
\end{align}
\begin{align}
\tilde e^{(9)}_{ij}(\omega, \vec{x}) =& -\frac{m}{8\pi l^2} \omega^2\int dt \frac{dt}{d\sigma} e^{i\omega t} v_i v_j \tilde{G}_l\left(\omega,\vec{x},\vec{r}\right)\,,
\end{align}
\begin{align}
\tilde e^{(10)}_{ij}(\omega, \vec{x}) =& \frac{m}{8\pi l^2}\int dt \frac{dt}{d\sigma} \frac{5}{8}e^{i\omega t}(v_i\n_j+v_j\n_i)\tilde{G}_l\left(\omega,\vec{x},\vec{r}\right)\,,
\end{align}
$\tilde e^{(1)}_{ij}(\omega, \vec{x})$ to $\tilde e^{(4)}_{ij}(\omega, \vec{x})$ are contributions of the black hole mass. $\tilde e^{(5)}_{ij}(\omega, \vec{x})$ to $\tilde e^{(7)}_{ij}(\omega, \vec{x})$ arises due to the charge of the black hole. $\tilde e^{(8)}_{ij}(\omega, \vec{x})$ to $\tilde e^{(10)}_{ij}(\omega, \vec{x})$ can be treated as the contribution from the AdS potential.

The contribution of black hole mass and charge to the soft factor for gravitation was explicitly derived in \cite{Fernandes:2020tsq}  
\begin{align}
S_{\text{gr}}^{\text{flat}} &= i \frac{4 \pi R}{e^{i \omega R}} \epsilon^{ij}\tilde{e}_{ij} (\omega, \vec{x})\notag\\
&= - \frac{m}{\omega} \epsilon^{ij}\left[\frac{1}{1 - \hat{n}.\vec{\beta}_+}\frac{1}{\sqrt{1 - \vec{\beta}^2_+}}\beta_{+i} \beta_{+j} - \frac{1}{1 - \hat{n}.\vec{\beta}_-}\frac{1}{\sqrt{1 - \vec{\beta}^2_-}}\beta_{-i} \beta_{-j}\right] \notag\\
&\quad -im\ln \omega^{-1}\epsilon^{ij}\,\left[{1\over \sqrt{1-\vec \beta_+^2}}\, \left\{C_+ \left({1\over 1-\hat n.\vec\beta_+} +{1\over 1 -\vec\beta_+^2}\right)-{ M\over 8\, \pi\, |\vec\beta_+|^3}\, {3\vec\beta_+^2 -1 \over 1-\vec \beta_+^2}\right\}\beta_{+ i} 
\beta_{+ j}  \right.\nonumber \\ &
\left. - {1\over \sqrt{1-\vec \beta_-^2}}\,  \left\{C_- \left({1\over 1-\hat n.\vec\beta_-}+{1\over 1 -\vec\beta_-^2} \right) + 
{M\over 8\, \pi\, |\vec\beta_-|^3} {3\vec\beta_-^2-1\over 1-\vec\beta_-^2}
\right\}\beta_{- i} \beta_{- j} \right] \notag\\
& \qquad - i m \frac{M}{4 \pi} \ln \left(R \omega\right) \epsilon^{ij}\left[\frac{1}{1 - \hat{n}.\vec{\beta}_+}\frac{1}{\sqrt{1 - \vec{\beta}^2_+}}\beta_{+i} \beta_{+j} - \frac{1}{1 - \hat{n}.\vec{\beta}_-}\frac{1}{\sqrt{1 - \vec{\beta}^2_-}}\beta_{-i} \beta_{-j}\right] \notag\\
& \quad \qquad -i \frac{qQ}{4\pi}\ln \omega^{-1}\epsilon^{ij}\,\left[ \frac{\beta_{+ i}\beta_{+ j}}{ |\vec\beta_+|^3}+\frac{\beta_{- i}\beta_{- j}}{ |\vec\beta_-|^3}\right] + \text{finite}\,.\label{eij.soft1}
\end{align}
 To complete the analysis we need the expression for $C_{\pm}$, which can be determined from considering the energy conservation equation. The energy of the probe particle can be written from the point particle action in \ref{act.pp} as
\begin{align}
E =m\vert g_{00}\vert \frac{dt}{d\sigma}-\frac{q}{4\pi}A_0\,.
\label{eom.pp}
\end{align}
Expanding this expression in powers of $t$, we then find from the $\frac{1}{t}$ coefficient the following relation of $C_{\pm}$ with $M$ and $Q$ \cite{Fernandes:2020tsq} 
\begin{align}
C_{\pm}=\mp{M\over 8\, \pi\, |\vec\beta_{\pm}|^3} (1-3\vec\beta_{\pm}^2)\mp\frac{qQ}{4\pi m  |\vec\beta_{\pm}|^3}(1-\vec\beta_{\pm}^2)^{3/2}\,.
\label{c1.pp}
\end{align}
Substituting for $M$ and $Q$ in \ref{eij.soft1} using \ref{c1.pp} gives
\begin{align} 
S_{\text{gr}}^{\text{flat}} =& - \frac{m}{\omega} \epsilon^{ij}\left[\frac{1}{1 - \hat{n}.\vec{\beta}_+}\frac{1}{\sqrt{1 - \vec{\beta}^2_+}}\beta_{+i} \beta_{+j} - \frac{1}{1 - \hat{n}.\vec{\beta}_-}\frac{1}{\sqrt{1 - \vec{\beta}^2_-}}\beta_{-i} \beta_{-j}\right] \notag\\
&-i m \ln \omega^{-1} \epsilon^{ij} \left[C_+\frac{1}{1 - \hat{n}.\vec{\beta}_+}\frac{1}{\sqrt{1 - \vec{\beta}^2_+}}\beta_{+i} \beta_{+j} - C_-\frac{1}{1 - \hat{n}.\vec{\beta}_-}\frac{1}{\sqrt{1 - \vec{\beta}^2_-}}\beta_{-i} \beta_{-j}\right] \notag\\
&- i m \frac{M}{4 \pi} \ln \left(R \omega\right) \epsilon^{ij}\left[\frac{1}{1 - \hat{n}.\vec{\beta}_+}\frac{1}{\sqrt{1 - \vec{\beta}^2_+}}\beta_{+i} \beta_{+j} - \frac{1}{1 - \hat{n}.\vec{\beta}_-}\frac{1}{\sqrt{1 - \vec{\beta}^2_-}}\beta_{-i} \beta_{-j}\right]\,.
\label{eij.soft}
\end{align}
This result agrees with the classical limit of the soft graviton factor, up to the subleading logarithmic contribution, in asymptotically flat spacetimes. 
In \cite{Banerjee:2020dww}, the leading order contribution of cosmological constant to the soft factor was derived with the following expression
\begin{align}
& S_{\rm gr}^l=\notag\\&-\frac{m}{2\gamma^2}\, \omega^{-1}\, \epsilon^{ij} \left\{\frac{1}{(1- \hat{n}.\vec{\beta})^3}\beta_{+ i} \beta_{+ j}\frac{\vec{\beta}_{+}^2 (3-2\vec{\beta}_{+}^2)}{(1-\vec{\beta}_{+}^2)^{\frac{3}{2}}} - \frac{1}{(1- \hat{n}.\vec{\beta})^3}\beta_{- i} \beta_{- j}\frac{\vec{\beta}_{-}^2 (3-2\vec{\beta}_{-}^2)}{(1-\vec{\beta}_{-}^\frac{3}{2})}\right\}\notag\\& - i\, \frac{m}{2\gamma^2} \,   \ln \omega^{-1} \, \epsilon^{ij}\, \left[\frac{\beta_{+ i} \beta_{+ j}}{(1- \hat{n}.\vec{\beta}_+)^3}\frac{C_+ \vec{\beta}_+^2(3-2\vec{\beta}_+^2)}{\left(1-\vec{\beta}_+^2\right)^{\frac{3}{2}}} - \frac{\beta_{- i} \beta_{- j}}{(1- \hat{n}.\vec{\beta}_-)^3}\frac{C_- \vec{\beta}_-^2(3-2\vec{\beta}_-^2)}{\left(1-\vec{\beta}_-^2\right)^{\frac{3}{2}}}\right]\,.\label{fr}
\end{align}


\begin{thebibliography}{100}

\bibitem{Weinberg:1964ew}
S.~Weinberg,
Phys. Rev. \textbf{135}, B1049-B1056 (1964)
doi:10.1103/PhysRev.135.B1049

\bibitem{Weinberg:1965nx}
S.~Weinberg,
Phys. Rev. \textbf{140}, B516-B524 (1965)
doi:10.1103/PhysRev.140.B516

\bibitem{Kapec:2014zla}
D.~Kapec, V.~Lysov and A.~Strominger,
Adv. Theor. Math. Phys. \textbf{21}, 1747-1767 (2017)
doi:10.4310/ATMP.2017.v21.n7.a6
[arXiv:1412.2763 [hep-th]].

\bibitem{He:2014cra}
T.~He, P.~Mitra, A.~P.~Porfyriadis and A.~Strominger,
JHEP \textbf{10}, 112 (2014)
doi:10.1007/JHEP10(2014)112
[arXiv:1407.3789 [hep-th]].

\bibitem{Strominger:2013jfa}
A.~Strominger,
JHEP \textbf{07}, 152 (2014)
doi:10.1007/JHEP07(2014)152
[arXiv:1312.2229 [hep-th]].

\bibitem{Cachazo:2014fwa}
F.~Cachazo and A.~Strominger,
[arXiv:1404.4091 [hep-th]].

\bibitem{Campiglia:2015yka}
M.~Campiglia and A.~Laddha,
JHEP \textbf{04}, 076 (2015)
doi:10.1007/JHEP04(2015)076
[arXiv:1502.02318 [hep-th]].

\bibitem{Lysov:2014csa}
V.~Lysov, S.~Pasterski and A.~Strominger,
Phys. Rev. Lett. \textbf{113}, no.11, 111601 (2014)
doi:10.1103/PhysRevLett.113.111601
[arXiv:1407.3814 [hep-th]].

\bibitem{Schwab:2014xua}
B.~U.~W.~Schwab and A.~Volovich,
Phys. Rev. Lett. \textbf{113}, no.10, 101601 (2014)
doi:10.1103/PhysRevLett.113.101601
[arXiv:1404.7749 [hep-th]].

\bibitem{Campiglia:2014yka}
M.~Campiglia and A.~Laddha,
Phys. Rev. D \textbf{90}, no.12, 124028 (2014)
doi:10.1103/PhysRevD.90.124028
[arXiv:1408.2228 [hep-th]].

\bibitem{Casali:2014xpa}
E.~Casali,
JHEP \textbf{08}, 077 (2014)
doi:10.1007/JHEP08(2014)077
[arXiv:1404.5551 [hep-th]].

\bibitem{Broedel:2014fsa}
J.~Broedel, M.~de Leeuw, J.~Plefka and M.~Rosso,
Phys. Rev. D \textbf{90}, no.6, 065024 (2014)
doi:10.1103/PhysRevD.90.065024
[arXiv:1406.6574 [hep-th]].

\bibitem{Conde:2016csj}
E.~Conde and P.~Mao,
Phys. Rev. D \textbf{95}, no.2, 021701 (2017)
doi:10.1103/PhysRevD.95.021701
[arXiv:1605.09731 [hep-th]].

\bibitem{Chakrabarti:2017zmh}
S.~Chakrabarti, S.~P.~Kashyap, B.~Sahoo, A.~Sen and M.~Verma,
JHEP \textbf{01}, 090 (2018)
doi:10.1007/JHEP01(2018)090
[arXiv:1709.07883 [hep-th]].

\bibitem{Chakrabarti:2017ltl}
S.~Chakrabarti, S.~P.~Kashyap, B.~Sahoo, A.~Sen and M.~Verma,
JHEP \textbf{12}, 150 (2017)
doi:10.1007/JHEP12(2017)150
[arXiv:1707.06803 [hep-th]].

\bibitem{Laddha:2017vfh}
A.~Laddha and P.~Mitra,
JHEP \textbf{05}, 132 (2018)
doi:10.1007/JHEP05(2018)132
[arXiv:1709.03850 [hep-th]].

\bibitem{AtulBhatkar:2018kfi}
S.~Atul Bhatkar and B.~Sahoo,
JHEP \textbf{01}, 153 (2019)
doi:10.1007/JHEP01(2019)153
[arXiv:1809.01675 [hep-th]].

\bibitem{Addazi:2019mjh}
A.~Addazi, M.~Bianchi and G.~Veneziano,
JHEP \textbf{05}, 050 (2019)
doi:10.1007/JHEP05(2019)050
[arXiv:1901.10986 [hep-th]].

\bibitem{Sahoo:2020ryf}
B.~Sahoo,
JHEP \textbf{12}, 070 (2020)
doi:10.1007/JHEP12(2020)070
[arXiv:2008.04376 [hep-th]].

\bibitem{Sahoo:2018lxl}
B.~Sahoo and A.~Sen,
JHEP \textbf{02}, 086 (2019)
doi:10.1007/JHEP02(2019)086
[arXiv:1808.03288 [hep-th]].

\bibitem{Laddha:2018rle}
A.~Laddha and A.~Sen,
JHEP \textbf{09}, 105 (2018)
doi:10.1007/JHEP09(2018)105
[arXiv:1801.07719 [hep-th]].

\bibitem{Laddha:2018myi}
A.~Laddha and A.~Sen,
JHEP \textbf{10}, 056 (2018)
doi:10.1007/JHEP10(2018)056
[arXiv:1804.09193 [hep-th]].

\bibitem{Laddha:2018vbn}
A.~Laddha and A.~Sen,
Phys. Rev. D \textbf{100}, no.2, 024009 (2019)
doi:10.1103/PhysRevD.100.024009
[arXiv:1806.01872 [hep-th]].

\bibitem{Laddha:2019yaj}
A.~Laddha and A.~Sen,
Phys. Rev. D \textbf{101}, no.8, 084011 (2020)
doi:10.1103/PhysRevD.101.084011
[arXiv:1906.08288 [gr-qc]].

\bibitem{Saha:2019tub}
A.~P.~Saha, B.~Sahoo and A.~Sen,
JHEP \textbf{06}, 153 (2020)
doi:10.1007/JHEP06(2020)153
[arXiv:1912.06413 [hep-th]].

\bibitem{Gary:2009mi}
M.~Gary and S.~B.~Giddings,
Phys. Rev. D \textbf{80}, 046008 (2009)
doi:10.1103/PhysRevD.80.046008
[arXiv:0904.3544 [hep-th]].

\bibitem{Penedones:2010ue}
J.~Penedones,
JHEP \textbf{03}, 025 (2011)
doi:10.1007/JHEP03(2011)025
[arXiv:1011.1485 [hep-th]].

\bibitem{Fitzpatrick:2011ia}
A.~L.~Fitzpatrick, J.~Kaplan, J.~Penedones, S.~Raju and B.~C.~van Rees,
JHEP \textbf{11}, 095 (2011)
doi:10.1007/JHEP11(2011)095
[arXiv:1107.1499 [hep-th]].

\bibitem{Rastelli:2016nze}
L.~Rastelli and X.~Zhou,
Phys. Rev. Lett. \textbf{118}, no.9, 091602 (2017)
doi:10.1103/PhysRevLett.118.091602
[arXiv:1608.06624 [hep-th]].

\bibitem{Hijano:2020szl}
E.~Hijano and D.~Neuenfeld,
JHEP \textbf{11}, 009 (2020)
doi:10.1007/JHEP11(2020)009
[arXiv:2005.03667 [hep-th]].

\bibitem{Compere:2019bua}
G.~Comp\`ere, A.~Fiorucci and R.~Ruzziconi,
Class. Quant. Grav. \textbf{36} (2019) no.19, 195017
doi:10.1088/1361-6382/ab3d4b
[arXiv:1905.00971 [gr-qc]].

\bibitem{Fernandes:2020tsq}
K.~Fernandes and A.~Mitra,
Phys. Rev. D \textbf{102}, no.10, 105015 (2020)
doi:10.1103/PhysRevD.102.105015
[arXiv:2005.03613 [hep-th]].

\bibitem{Banerjee:2020dww}
N.~Banerjee, A.~Bhattacharjee and A.~Mitra,
JHEP \textbf{01}, 038 (2021)
doi:10.1007/JHEP01(2021)038
[arXiv:2008.02828 [hep-th]].

\bibitem{He:2014laa}
T.~He, V.~Lysov, P.~Mitra and A.~Strominger,
JHEP \textbf{05}, 151 (2015)
doi:10.1007/JHEP05(2015)151
[arXiv:1401.7026 [hep-th]].

\bibitem{Hamada:2018vrw}
Y.~Hamada and G.~Shiu,
Phys. Rev. Lett. \textbf{120}, no.20, 201601 (2018)
doi:10.1103/PhysRevLett.120.201601
[arXiv:1801.05528 [hep-th]].

\bibitem{Campiglia:2019wxe}
M.~Campiglia and A.~Laddha,
JHEP \textbf{10}, 287 (2019)
doi:10.1007/JHEP10(2019)287
[arXiv:1903.09133 [hep-th]].

\bibitem{AtulBhatkar:2019vcb}
S.~Atul Bhatkar,
JHEP \textbf{10}, 110 (2020)
doi:10.1007/JHEP10(2020)110
[arXiv:1912.10229 [hep-th]].

\bibitem{Wald:1984rg}
R.~M.~Wald,
doi:10.7208/chicago/9780226870373.001.0001

\bibitem{DeWitt:1960fc}
B.~S.~DeWitt and R.~W.~Brehme,
Annals Phys. \textbf{9}, 220-259 (1960)
doi:10.1016/0003-4916(60)90030-0

\bibitem{Peters:1966} 
  P.~C.~Peters,
  ``Perturbations in the Schwarzschild Metric,''
  Phys.\ Rev.\ {\bf146}, 938 (1966). 
  doi:10.1103/PhysRev.146.938
  
\bibitem{Peters:1970mx}
P.~C.~Peters,
Phys. Rev. D \textbf{1}, 1559-1571 (1970)
doi:10.1103/PhysRevD.1.1559

\bibitem{Kovacs:1977uw}
S.~J.~Kovacs and K.~S.~Thorne,
Astrophys. J. \textbf{217}, 252-280 (1977)
doi:10.1086/155576

\bibitem{Poisson:2011nh}
E.~Poisson, A.~Pound and I.~Vega,
Living Rev. Rel. \textbf{14}, 7 (2011)
doi:10.12942/lrr-2011-7
[arXiv:1102.0529 [gr-qc]].

\bibitem{Strominger:2017zoo}
A.~Strominger,
[arXiv:1703.05448 [hep-th]].

\bibitem{Hijano:2019qmi}
E.~Hijano,
JHEP \textbf{07}, 132 (2019)
doi:10.1007/JHEP07(2019)132
[arXiv:1905.02729 [hep-th]].

\bibitem{Hamilton:2006az}
A.~Hamilton, D.~N.~Kabat, G.~Lifschytz and D.~A.~Lowe,
Phys. Rev. D \textbf{74}, 066009 (2006)
doi:10.1103/PhysRevD.74.066009
[arXiv:hep-th/0606141 [hep-th]].
\end{thebibliography}
\end{document}